\documentclass[aps,prb,twocolumn,floatfix,superscriptaddress,longbibliography]{revtex4-2}

\usepackage{physics}
\usepackage{float}
\usepackage{bbold}
\usepackage{color}
\usepackage[english]{babel}
\usepackage[utf8]{inputenc}
\usepackage{amssymb}
\usepackage{amsmath}
\usepackage{dsfont}
\usepackage{multirow,tabularx,makecell}
\usepackage[pdftex]{graphicx}
\usepackage{xspace}
\usepackage{dsfont}
\usepackage{bm}
\usepackage{nicefrac}
\usepackage[export]{adjustbox}
\usepackage[pdfencoding=auto, psdextra]{hyperref}
\hypersetup{
    colorlinks,%
    citecolor=blue,%
    filecolor=blue,%
    linkcolor=blue,%
    urlcolor=blue
}

\usepackage[usenames, dvipsnames]{xcolor}

\begin{document}

\title{Skyrmion motion in magnetic anisotropy gradients: Acceleration caused by deformation}

\author{Ismael Ribeiro de Assis}
\affiliation{Institut f\"ur Physik, Martin-Luther-Universit\"at Halle-Wittenberg, D-06099 Halle (Saale), Germany}

\author{Ingrid Mertig}
\affiliation{Institut f\"ur Physik, Martin-Luther-Universit\"at Halle-Wittenberg, D-06099 Halle (Saale), Germany}

\author{B{\"o}rge G{\"o}bel}

\affiliation{Institut f\"ur Physik, Martin-Luther-Universit\"at Halle-Wittenberg, D-06099 Halle (Saale), Germany}

\date{\today}

\begin{abstract}
Magnetic skyrmions are nano-sized topologically non-trivial spin textures that can be moved by external stimuli such as spin currents and internal stimuli such as spatial gradients of a material parameter. Since the total energy of a skyrmion depends linearly on most of these parameters, like the perpendicular magnetic anisotropy, the exchange constant, or the Dzyaloshinskii–Moriya interaction strength, a skyrmion will move uniformly in a weak parameter gradient. In this paper, we show that the linear behavior changes once the gradients are strong enough so that the magnetic profile of a skyrmion is significantly altered throughout the propagation. In that case, the skyrmion experiences acceleration and moves along a curved trajectory. Furthermore, we show that when spin-orbit torques and material parameter gradients trigger a skyrmion motion, it can move on a straight path along the current or gradient direction. We discuss the significance of suppressing the skyrmion Hall effect for spintronic and neuromorphic applications of skyrmions. Lastly, we extend our discussion and compare it to a gradient generated by the Dzyaloshinskii–Moriya interaction.
\end{abstract}

\maketitle

\section{Introduction}

\begin{figure*}[!t]
  \centering
    \includegraphics[width=\textwidth]{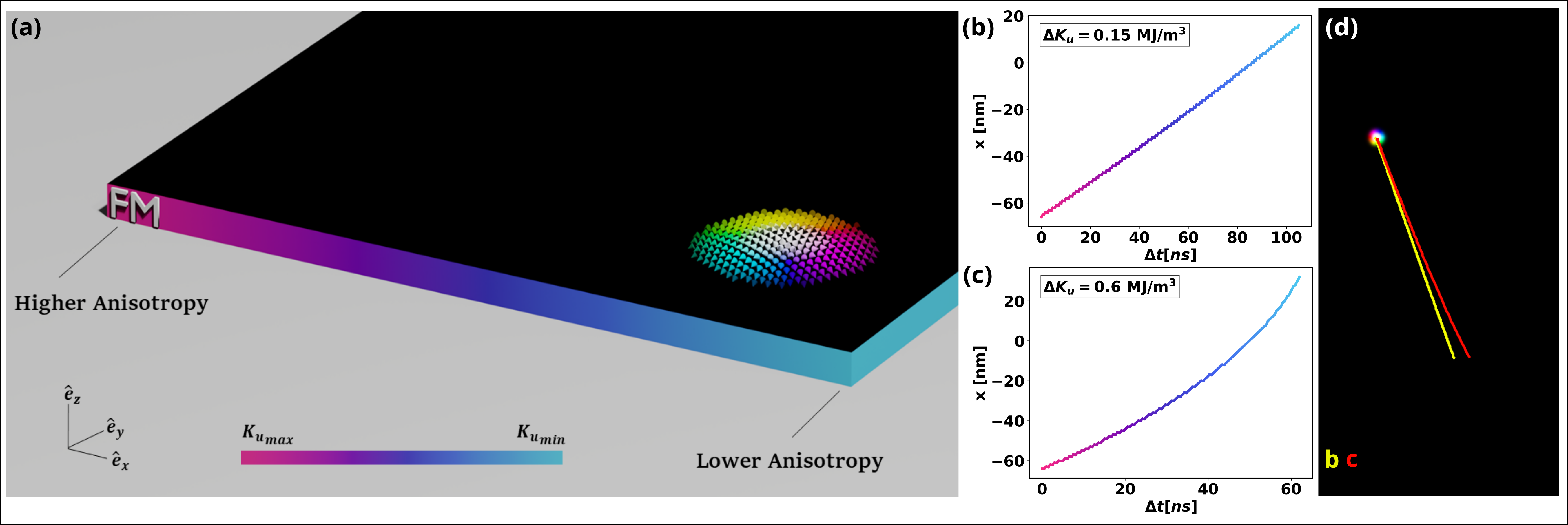}
    \caption{Skyrmion motion in a magnetic anisotropy gradient. (a) Schematic illustration of the gradient in the ferromagnetic layer (FM). The colored cones represent the orientation of the magnetic moments that form the skyrmion. (b,c) Change of the $x$ coordinate of the skyrmion plotted against time in two different anisotropy gradients. (b) is the result for the weakest gradient that we have analyzed. (c) is the result for the strongest gradient that we have analyzed. The color represents the corresponding value of $K_\mathrm{u}$ as shown in panel (a). (d) The trajectories of the skyrmion moving according to the two simulations are shown in (b) yellow and (c) red.}
    \label{fig1:FM_skyrmion}
\end{figure*}

Magnetic skyrmions~\cite{bogdanov1989thermodynamically, nagaosa2013topological} have emerged as promising candidates for various applications, including data storages~\cite{fert2013skyrmions} and neuromorphic computing~\cite{li2021magnetic, li2017magnetic,de2023biskyrmion}. These nano-objects are topological-protected spin textures that have attracted enormous research interest due to their high stability, compact size, and particle-like behavior. Several types of skyrmions can be stabilized in thin films~\cite{gobel2021beyond}, such as Néel~\cite{heinze2011spontaneous}, Bloch~\cite{muhlbauer2009skyrmion} and antiskyrmions~\cite{nayak2017magnetic}. Their size and profile result from the interplay of several interactions, such as exchange interaction, Dzyaloshinskii–Moriya interaction (DMI) and perpendicular magnetic anisotropy (PMA)~\cite{nagaosa2013topological,gobel2021beyond}.

Most of the envisioned skyrmion-based applications rely on a precisely controlled current-driven motion which is, however, compromised by the skyrmion Hall effect (SkHE)~\cite{jiang2017direct0,litzius2017skyrmion}: Skyrmions are subject to a gyroscopic force due to their non-trivial topology, which leads to a transverse velocity component and a motion towards the edge of the sample, potentially leading to skyrmion annihilation. Therefore, understanding and controlling their dynamics is crucial for harnessing their full potential for spintronic applications. 

One approach to circumvent the SkHE bottleneck is the skyrmion-driven motion by internal magnetic parameter gradients, most importantly the PMA and DMI gradients. The corresponding gradient in a skyrmion's energy induces the skyrmion's motion. Several approaches for gradient engineering have been demonstrated experimentally, including but not limited to magnetic patterning with helium ions~\cite{juge2021helium, kern2022deterministic,kern2022tailoring,ahrens2022skyrmion}, thickness variations~\cite{yu2016room} and inducing strain~\cite{gusev2020manipulation}. In parallel, further research \cite{tomasello2018chiral, gorshkov2022dmi} has investigated theoretically the dynamics of skyrmions subjected to these gradients.

This paper highlights the crucial role of variations in the skyrmion profile for accelerated motion in a PMA gradient. We study a Néel skyrmion characterized by the topological charge $N_\mathrm{sk} = +1$, with a comprehensive numerical analysis through micromagnetic simulations using the software package Mumax3~\cite{vansteenkiste2011mumax,vansteenkiste2014design}. While a skyrmion moves uniformly when the parameter gradient is so weak that the skyrmion profile remains unchanged, the skyrmion deforms in a strong gradient causing an accelerated type of motion along a curved trajectory. Furthermore, we present two possible scenarios to suppress the SkHE, each with distinct potential application. Lastly, we compare the PMA gradient with a DMI gradient. Notably, all discussions can also be applied to a skyrmion moving in DMI gradients with only minor differences. 

This paper is structured as follows: After explaining the theoretical framework of the micromagnetic simulations in Sec.~\ref{sec:modelmethods}, we discuss the simulated trajectories of skyrmions in various PMA gradients in Sec.~\ref{sec:pmagradient}. We examine the alterations in the skyrmion's profile during movement to derive expressions for its energy, driving force, velocity, and acceleration. Besides the acceleration, the deformation of a skyrmion in a PMA gradient also affects the magnitude of the skyrmion Hall angle, as analyzed in Sec.~\ref{sec:dissipative}, leading to curved trajectories. In Sec.~\ref{sec:suppression}, we simulate the motion of skyrmions in PMA gradients and under the influence of spin-orbit torques generated by electrical currents. We show that the SkHE induced by the current can compensate the SkHE caused by the PMA, leading to a suppressed transverse motion. As presented in Sec.~\ref{sec:comparison}, the results derived for the PMA gradient carry over to other parameter gradients, like gradients in the DMI strength, as well. We summarize and elaborate on the significance of our results in Sec.~\ref{sec:conclusion}.

\section{Model and Methods}\label{sec:modelmethods}

For all simulations described in this work, we have used the GPU-accelerated software package Mumax3~\cite{vansteenkiste2011mumax,vansteenkiste2014design} to solve the Landau-Lifshitz-Gilbert (LLG) equation~\cite{landau1992theory} for each unit vector $\bm{m}_i$ indicating the magnetization at a cell $i$ of the sample
\begin{equation}
\partial_t \bm{m}_i = -\gamma_e \bm{m}_i \times \bm{B}^{i}_\mathrm{eff} + \alpha  \bm{m}_i \times  \partial_t \bm{m}_i + \gamma_e \epsilon \beta[( \bm{m}_i \times \bm{s}) \times \bm{m}_i].\label{eq:LLG_mumax3}
\end{equation}
Besides the precession and Gilbert-damping term ($\alpha= 0.3$), simulations in the later part of the paper include the spin-orbit torque (SOT) term. Here a charge current is translated into a spin current by the spin Hall effect. The spins $\bm{s}$ enter the magnetic sample perpendicularly. 
The constants in Eq.~\eqref{eq:LLG_mumax3} are: the gyromagnetic ration $\gamma_e = 1.760 \times 10^{11} $ T$^{-1}$s$^{-1}$ and $\epsilon \beta= \frac{\hbar \Theta_\mathrm{SH}}{2 e d_z M_s}$; where $d_z$ is the thickness of the magnetic layer, $e$ the electron's charge, $\hbar$ Planck's constant, $M_s$ the saturation magnetization and $j\Theta_\mathrm{SH}$ the spin current with spin orientation $\bm{s}$ and the spin Hall angle $\Theta_\mathrm{SH}$.

$\bm{B}^{i}_\mathrm{eff} = \delta F/M_s \delta \bm{m}_i$ is the effective magnetic field derived from the system's total free energy density $F$. The total energy 
\begin{equation}
    E = E_{\mathrm{ex}} + E_{\mathrm{DMI}} + E_{\mathrm{anis}}
    \label{eq:Total_energy}
\end{equation}
is the sum of the exchange energy
\begin{equation}
E_\mathrm{ex}=\int A(\bm{r'}) |\nabla \bm{m}(\bm{r}')|^2 \mathrm{d}^3r',
\end{equation}
the magnetocrystalline anisotropy 
\begin{equation}
E_\mathrm{anis} = \int K_\mathrm{u}(\bm{r'})  [1 -  m_{z}(\bm{r}')^2 
]\mathrm{d}^3r',
\end{equation}
the interfacial DMI 
\begin{equation}
E_\mathrm{DMI}=\int D_i(\bm{r'}) [m_z(\bm{r'}) \nabla \cdot \bm{m}(\bm{r}') - (\bm{m}(\bm{r'}) \cdot \nabla) m_z(\bm{r'})]\mathrm{d}^3r'
\end{equation}
and the demagnetization field (dipole-dipole interaction). We simulate an interface of CoPt as in Ref.~\cite{sampaio2013nucleation} and use the magnetic parameters $A_\mathrm{ex}= 1.5 \times 10^{11}$ J/m and $D_\mathrm{i}= 3.5 \times 10^{-3}$ J/m$^2$. The anisotropy value is varied near the value $K_\mathrm{u}= 8.0 \times 10^5$ J/m$^3$ as explained in the following.

The ferromagnetic layer has dimensions of $252\,\mathrm{nm} \times 512\,\mathrm{nm} \times 1\,\mathrm{nm}$ in all simulations and a cell size of $1\,\mathrm{nm} \times 1\,\mathrm{nm} \times 1\,\mathrm{nm}$. To simulate the anisotropy gradient, we divide the FM layer into 252 regions, each having a different value of $K_\mathrm{u}$. In other words, each cell along the $x$ direction has a unique $K_\mathrm{u}$ parameter. The anisotropy varies linearly starting from $K_{u_\mathrm{max}}$ on the left side of the FM layer and decreasing to $K_{u_\mathrm{min}}$ on the right side, as schematically shown in Fig.~\ref{fig1:FM_skyrmion}(a).

\section{Results and Discussion}

We start by discussing the motion of skyrmions in PMA gradients, highlighting the relevance of deformations for an accelerated type of motion (Sec.~\ref{sec:pmagradient}). After that, we analyze the curved trajectories (Sec.~\ref{sec:dissipative}) and present how SOTs can suppress the SkHE caused by the PMA gradients (Sec.~\ref{sec:suppression}). Finally, we generalize our results to other magnetic parameter gradients (Sec.~\ref{sec:comparison}).

\subsection{Skyrmion motion in a PMA gradient}\label{sec:pmagradient}

In this section, we simulate a ferromagnetic layer (FM) in the $xy$ plane, with a fixed exchange constant $A\equiv A(\bm{r}')$ and interfacial DMI coefficient $D_i\equiv D_i(\bm{r}')$. The PMA $K_\mathrm{u}(\bm{r})$ forms the gradient varying linearly along the $x$ direction 
\begin{equation}
K_\mathrm{u}(x)=\frac{\Delta K_\mathrm{u}}{l}x
\end{equation}
with $ \Delta K_\mathrm{u} = K_{u_\mathrm{max}} - K_{u_\mathrm{min}}$ and the length of the sample $l=252\,\mathrm{nm}$. 

As long as the anisotropy parameter does not change strongly over the size of the skyrmion, we can approximate 
\begin{equation}
E_{\mathrm{anis}} \approx K_\mathrm{u}(x) \int [ 1 -  m_{z}(\bm{r}')^2 ]\mathrm{d}^3r'
\end{equation}
because the integrand is only different from zero inside of the skyrmion. Unless stated otherwise, we refer to $K_\mathrm{u}(x)$ as the anisotropy at the center of the skyrmion. This approximation is valid for all simulations presented in this paper. 

As long as the spin texture of the skyrmion does not deform under motion, i.\,e., $c\equiv\int [ 1 -  m_{z}(\bm{r}')^2 ]\mathrm{d}^3r'=\mathrm{const.}$, the anisotropy energy can be approximated as $E_{\mathrm{anis}}(x) \approx c\frac{\Delta K_\mathrm{u}}{l}x$. In other words, if the skyrmion does not change, the anisotropy energy changes linearly along the sample.
However, if the profile changes, $c\neq \mathrm{const.}$, the energy changes non-linearly, giving rise to a fundamentally different propagation mode that we will discuss in the following.

To investigate the gradient-driven motion of skyrmions, we start by carrying out two simulations with distinct PMA gradient strengths $ \Delta K_\mathrm{u} / l$. We write and stabilize a Néel-Skyrmion with topological charge $N_\mathrm{Sk}=+1$, and let it propagate according to the LLG equation~\eqref{eq:LLG_mumax3}. The resulting changes in the coordinate and the trajectories are shown in Fig.~\ref{fig1:FM_skyrmion}(b-d). In the following, we refer to these simulations as simulation 1 (weak gradient) with $\Delta K_\mathrm{u} = 0.15$ MJ/m$^3$ and  simulation 2 (strong gradient) with  $\Delta K_\mathrm{u} = 0.6$ MJ/m$^3$. The maximum $K_\mathrm{u}$ is fixed for both simulations with $K_{u_\mathrm{max}}= 1.2$ MJ/m$^3$, and we vary only $ K_{u_\mathrm{min}} = 1.05$ MJ/m$^3$ for simulation 1 and $ K_{u_\mathrm{min}} = 0.6$ MJ/m$^3$ for simulation 2.

In both cases, the skyrmion moves from a higher to a lower PMA region to minimize the total energy.
However, it does not follow the gradient but instead moves along the so-called skyrmion Hall angle $\theta_\mathrm{SkHE} = v_y/v_x$, which varies slightly between the two simulations. $v_x$ and $ v_y$ are the components of the velocity $\bm{v}$ along the $x$ and $y$ directions. 

\subsubsection{Uniform skyrmion motion in weak PMA gradients}

As we will elaborate in the following, the dynamics caused by the weak PMA gradient (simulation 1) is similar to that caused by a SOT: The skyrmion moves partially along the stimulus direction (gradient or applied current) but experiences a transverse deflection.

Assuming a rigid skyrmion profile, its dynamics can be accurately described by the Thiele equation
\begin{equation}\label{eq:Thiele}
    b \underline{\bm{G}} \times \bm{v} - b \underline{\bm{D}} \alpha \bm{v} - \nabla E(x) = 0,     
\end{equation}
which consists of three forces terms. The first term is the so-called gyroscopic force, characterized by the gyroscopic vector $\underline{\bm{G}} = 4 \pi N_\mathrm{Sk} \hat{\bm{e}}_z$, arising from the topological charge of the skyrmion $N_\mathrm{Sk} = \frac{1}{4 \pi} \int \bm{m} \cdot (\partial_x \bm{m} \times \partial_y \bm{m} ) \, \mathrm{d}^2 r$. The second term is the dissipative force, quantified by the Gilbert damping $\alpha$. The dissipative tensor 
$D_{ij} = \int ( \partial_{x_i} \bm{m} \cdot \partial_{x_j} \bm{m} ) \, \mathrm{d}^2 r$
only has non-zero $D_{xx}=D_{yy}\equiv D_0$ elements. Lastly, the third term is a force arising from the potential difference created by interactions with the edge, other skyrmions, or, as in our case, energy changes caused by the PMA gradient. The constant $b = M_s d_z / \gamma_e$ is determined from the sample parameters.

Since the system's total energy depends only on the skyrmion's $x$ position, the corresponding force is always along $x$. If the skyrmion does not deform, i.\,e. $c\equiv\int [ 1 -  m_{z}(\bm{r}')^2 ]\mathrm{d}^3r'=\mathrm{const.}$, it reads
\begin{equation}
-\nabla E(x) \approx -\nabla E_{\mathrm{anis}}(x) \approx -c\frac{\Delta K_\mathrm{u}}{l}\bm{e}_x.
\end{equation}
Note that all terms in the Thiele equation have the unit of a force, but since there is no mass term, they do not act in a Newtonian way. Instead, if all forces are constant, a constant velocity $\bm{v}$ can be calculated directly from the Thiele equation
\begin{align}
    v_x &= - \frac{D_0 \alpha}{b [(4\pi N_\mathrm{Sk})^2 + D_0^2 \alpha^2]}\cdot c\frac{\Delta K_\mathrm{u}}{l},
    \\
    v_y & = \frac{4\pi N_\mathrm{Sk}}{b [(4\pi N_\mathrm{Sk})^2 + D_0^2 \alpha^2]}\cdot c\frac{\Delta K_\mathrm{u}}{l},
    \label{eq:velocities}
\end{align}
quantifying the skyrmion Hall angle as
\begin{align}
    \tan\theta_\mathrm{SkHE} = \frac{v_y}{v_x}=\frac{4\pi N_\mathrm{Sk}}{\alpha D_{0}}.
\end{align}

The explained scenario of a constant force is a good approximation in the weak PMA gradient [simulation 1 presented in Fig.~\ref{fig1:FM_skyrmion}(b)]. The resulting trajectory is a straight line, and both coordinates increase linearly in time. Not surprisingly, this result is analogous to the propagation of a Néel skyrmion driven by SOT. More details on the motion caused by SOT follow in Sec.~\ref{sec:suppression}.


\subsubsection{Accelerated skyrmion motion in strong PMA gradients}

Figure.~\ref{fig1:FM_skyrmion}(c) shows the skyrmion propagation in the strong PMA gradient (simulation 2). In this case, the time evolution of the position is non-linear, indicating an acceleration of the skyrmion. Moreover, the direction of motion has changed slightly. A skyrmion Hall effect is still present, but the skyrmion Hall angle changes slightly during the propagation. As derived, for a rigid skyrmion profile, it should be constant, $\theta_\mathrm{SkHE} = \arctan\frac{4\pi N_\mathrm{Sk}}{\alpha D_{0}}$. However, $D_{0}$ changes during the propagation. 

Both effects, the continuous change of propagation velocity and direction, are caused by deformations of the skyrmion. We continue our discussion by analyzing the skyrmion profile to better understand and quantify these effects.

\begin{figure*}[]
    \centering
    \includegraphics[width=\textwidth]{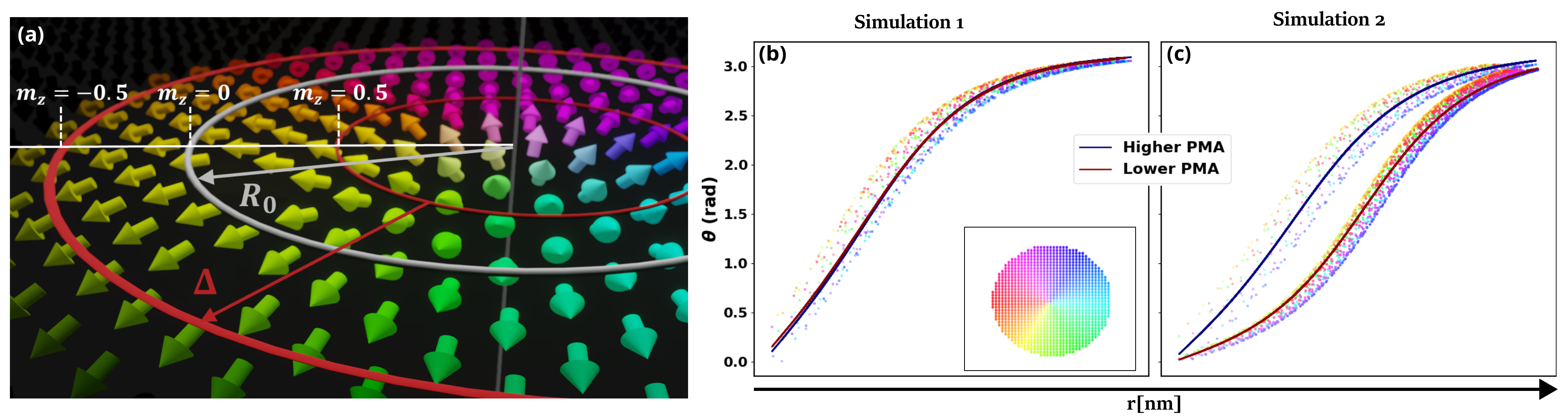}
    \caption{Characterization of the skyrmion profile. (a) Schematic figure of a skyrmion (formed by the colored arrows). The circles are lines of equal out-of-plane magnetization $m_z=-0.5, 0, +0.5$ as indicated. Their radii define the skyrmion's core radius $R_0$ and the domain wall width $\Delta$. (b,c) Skyrmion profile characterized by the azimuthal angle $\theta(r)$ at two different time steps in the simulation. Blue shows the earlier snapshot at $t=0$, corresponding to the higher anisotropy value. Red corresponds to the profile where the skyrmion has moved to the smaller anisotropy region, with $t= 93.5\,\mathrm{ns}$ for (b) and $t= 11.4\,\mathrm{ns}$ for (c). The lines are fits of the data points according to Eq.~\eqref{eq:Angle_skyrmion_profile}. The color of the points corresponds to the radial direction of the considered cut through the center of the skyrmion (cf. inset). The main text explains that the skyrmion is not perfectly rotational symmetric due to deformations caused by the anisotropy gradient, leading to scattered points.}
    \label{fig2:Skyrmion_profile}
\end{figure*}

\subsubsection{Deformation of the skyrmion profile}

In an ideal FM with homogeneous PMA (no gradient), the profile of the Néel skyrmion with $N_\mathrm{sk}=+1$ is rotational symmetric
\begin{align}
    m_x &= \frac{x}{r}\sin\theta\\
    m_y &= \frac{y}{r}\sin\theta\\
    m_z &= \cos\theta
\end{align}
and can thus be characterized by the azimuthal angle~\cite{romming2015field,buttner2018theory,wang2018theory}
\begin{equation}
\begin{aligned}
\theta(r) = 2 \arctan(e^{\frac{r + R_0}{\Delta}}) + 2 \arctan(e^{\frac{r - R_0}{\Delta}}) - \pi,
\end{aligned}
\label{eq:Angle_skyrmion_profile}
\end{equation}
where $r=\sqrt{x^2+y^2}$ is the distance from the skyrmion's center. $R_0$ is the radius of the skyrmion core and $\Delta$ the domain wall width. As depicted in Fig.~\ref{fig2:Skyrmion_profile}(a), $R_0$ is defined as the radius of the contour where $m_z=0$, and  $\Delta$ is defined by the width of the region that is enclosed by the contours for which $ -0.5 \le m_z \le 0.5$. 

$R_0$ and $\Delta$ allow us to quantify the deformation of the skyrmion during the propagation and help us to explain the observed acceleration in the strong PMA gradient.
In Fig.~\ref{fig2:Skyrmion_profile}(b,c), we plot the skyrmion profile at two distinct times in simulations 1 (b) and 2 (c), respectively. The time steps are such that the skyrmion is in a lower (red) and higher (blue) PMA region, respectively. Since the skyrmion is confined in the FM layer, which exhibits a PMA gradient, the Néel skyrmion is not rotational symmetric anymore. This is why we have determined the azimuthal angle profile $\theta(r)$ along various directions. The color of the data points corresponds to the different radial directions (see inset for explanation). Finally, the red and blue curves have been determined by fitting the data points according to Eq.~\eqref{eq:Angle_skyrmion_profile}. 

Note that to be able to use Eq.~\eqref{eq:Angle_skyrmion_profile}, we assume the contours $m_z =0$ and  $ -0.5 \le m_z \le 0.5$ in Fig.~\ref{fig2:Skyrmion_profile}(a) to be perfect circles. We attribute the non-perfect fit in Fig.~\ref{fig2:Skyrmion_profile} to this assumption, mainly to the fact that $\Delta$ is not the same on opposite sides of the skyrmion. Moreover, for small skyrmions, it becomes a numerical challenge to determine it. Nonetheless, Eq.~\eqref{eq:Angle_skyrmion_profile} can still provide a good description for the average change of $m_z$, and we shall use it to calculate $E_\mathrm{anis}$ later on. 

The presented procedure allows us to determine the skyrmion radius $R_0$ and the domain wall width $\Delta$ that define the skyrmion profile $\theta(r)$ for the two simulations at two different time steps, respectively. As expected from the trajectories presented in Fig.~\ref{fig1:FM_skyrmion}, the two skyrmion profiles are almost identical in the case of the weak gradient [red and blue overlap in Fig.~\ref{fig2:Skyrmion_profile}(b)]. This means there is almost no deformation. $R_0$ and $\Delta$ change equally by $ \approx 8\%$ comparing the start and the end of this simulation. However, for the strong PMA gradient, the skyrmion profile has changed significantly [Fig.~\ref{fig2:Skyrmion_profile}(c)]. Here, the skyrmion radius $R_0$ changes by $ \approx 133\%$ and the domain wall width $\Delta$ by $\approx 43\%$, after a simulation time of $11.4$ ns and a displacement $\Delta x \approx 92 $ nm  from its initial position $x_0 = -60$ nm. This severe change in the skyrmion profile leads to $c=c(x)\neq\mathrm{const.}$ and ultimately changes the energy landscape, giving rise to more complicated force terms that cause the skyrmion to accelerate. The anisotropy energy becomes
\begin{equation}
E_{\mathrm{anis}} \approx K_\mathrm{u}(x) \int [ 1 -  m_{z}(\bm{r}')^2 ]\mathrm{d}^3r'\approx c(x)\frac{\Delta K_\mathrm{u}}{l}x,
\end{equation}
which is a non-linear function in $x$, and the other energy terms become non-constant due to the deformation.

Before analyzing the energy terms in detail, we want to explain why the skyrmion deforms. As the skyrmion moves to the region with lower $K_\mathrm{u}$, the weaker anisotropy allows other competing interactions, such as the DMI and the exchange interaction, to become more influential in determining the skyrmion's profile. In response to the altered energy landscape, the DMI causes more magnetic moments to rotate, increasing $\Delta$. At the same time, the exchange interaction, which favors the parallel alignment of neighboring magnetic moments, comes into play. To maintain the energetically favorable alignment, the exchange interaction expands $R_0$. The next section analyzes how these profile changes affect the skyrmion's energy.

\subsubsection{Changes in the anisotropy energy}

Our goal is to express the anisotropy energy $E_\mathrm{anis}$ in terms of $x$ (and therefore $K_\mathrm{u}$). However, due to the change in the skyrmion profile, we have to take into account not only the explicit dependency via the term $K_\mathrm{u}(x)$ but also the implicit dependence via $c(x)$. Therefore, we express $E_\mathrm{anis}$ in terms of $R_0$ and $\Delta$, which we can easily fit with respect to $K_\mathrm{u}$ and $x$. In Ref.~\cite{buttner2018theory} this yielded the result
\begin{align}
\label{eq:Eanis}
E_\mathrm{anis}(x) &= 2 \pi K_\mathrm{u} \Delta^2 d_z I_\mathrm{anis}(R_0/ \Delta) + E_\mathrm{anis}^{0},
\end{align}
where $d_z$ is the sample thickness and $E_\mathrm{anis}^{0}$ a constant that accounts for the size of the sample and the average $K_\mathrm{u}$ value.

 Since we want to determine the force $-\nabla E_\mathrm{anis}$ that acts on the skyrmion and enters the Thiele equation, we are only interested in the terms that change with the coordinates; therefore, we shall drop $E_\mathrm{anis}^{0}$ for the rest of this work. $I_\mathrm{anis}$ is a non-elementary function that depends on the ratio $R_0/ \Delta$. It can be evaluated numerically and behaves like a quadratic function in $K_\mathrm{u}$ in the considered parameter range. We find a good agreement of Eq.~\eqref{eq:Eanis} with the micromagnetic data. Further details regarding the complete form of $I_\mathrm{anis}(R_0/ \Delta)$ and the fitting of Eq.~\eqref{eq:Eanis} to the micromagnetic data can be found in the Appendix.

\begin{figure}[]
    \centering
    \includegraphics[width=\columnwidth]{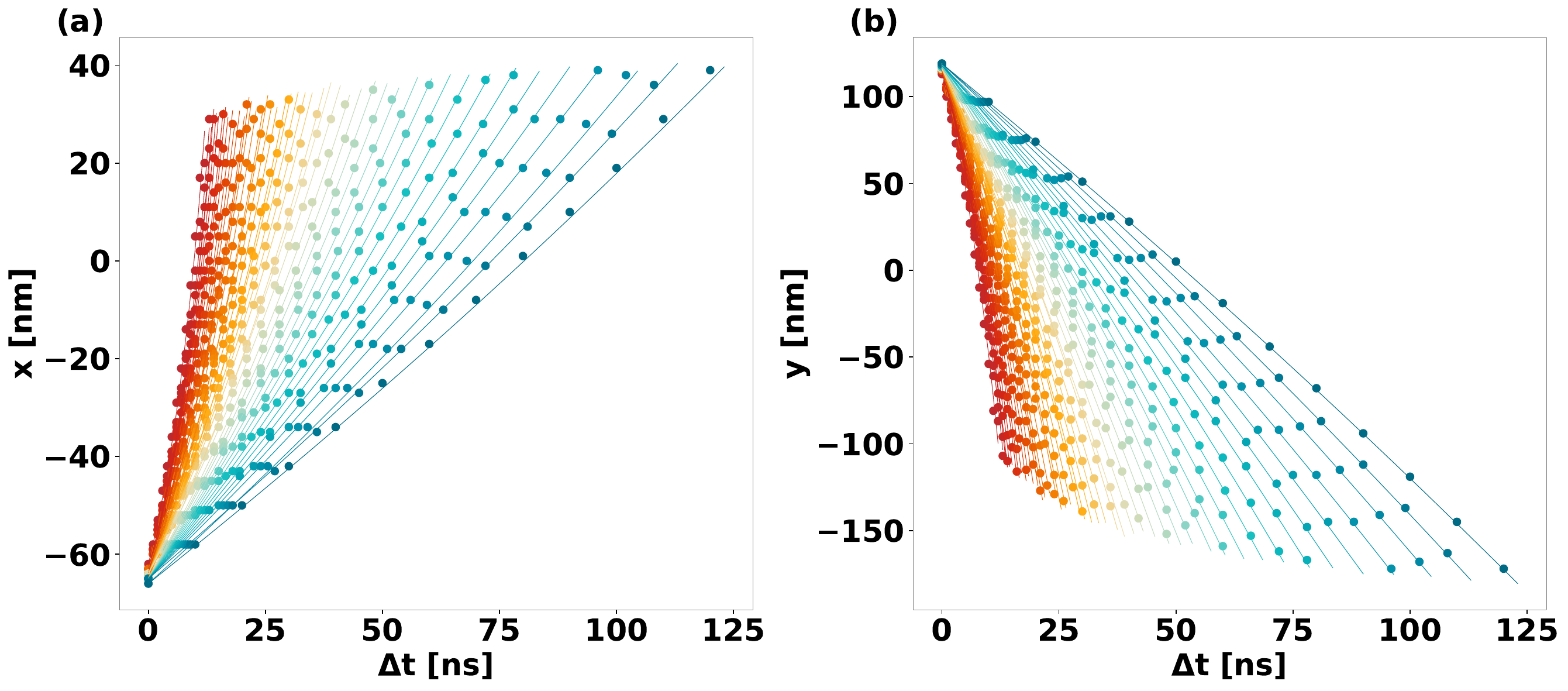}
    \caption{Motion of skyrmions in various PMA gradients. (a) The figure illustrates the $x$ coordinate of the skyrmions' centers over time. (b) shows the $y$ coordinate. The color represents the PMA gradient strength and is further explained in Fig. \ref{fig:quantities}. The data points are derived from micromagnetic simulations, while the overlaying lines correspond to the numerical solutions of the Thiele equation~\eqref{eq:Thiele}.}
    \label{fig:Skyrmion_trajectory}
\end{figure}

As we have seen in Fig.~\ref{fig2:Skyrmion_profile}, the skyrmion profile varies with the strength of $K_\mathrm{u}$ and therefore we can fit the dependencies $R_0(K_\mathrm{u})$ and $\Delta(K_\mathrm{u})$. For this purpose, we conduct 46 simulations with various PMA gradients. We maintained a constant $K_{\mathrm{u}_\mathrm{max}} = 1.2$ MJ/m$^3$ and varied $K_{\mathrm{u}_\mathrm{min}}$ in the distinct simulations: $0.6$ MJ/m$^3  \le K_{\mathrm{u}_\mathrm{min}} \le 1.05$ MJ/m$^3$. The limits are the two simulations discussed before. The trajectories are presented in Fig.~\ref{fig:Skyrmion_trajectory}. As expected, the array of curves falls between simulations 1 and 2. For weak PMA gradients, the coordinates change almost linearly, and the velocity is rather small. For strong PMA gradients, the dynamics become non-linear, and the velocity increases.

\begin{figure*}[]
  \centering
  \includegraphics[width=\textwidth]{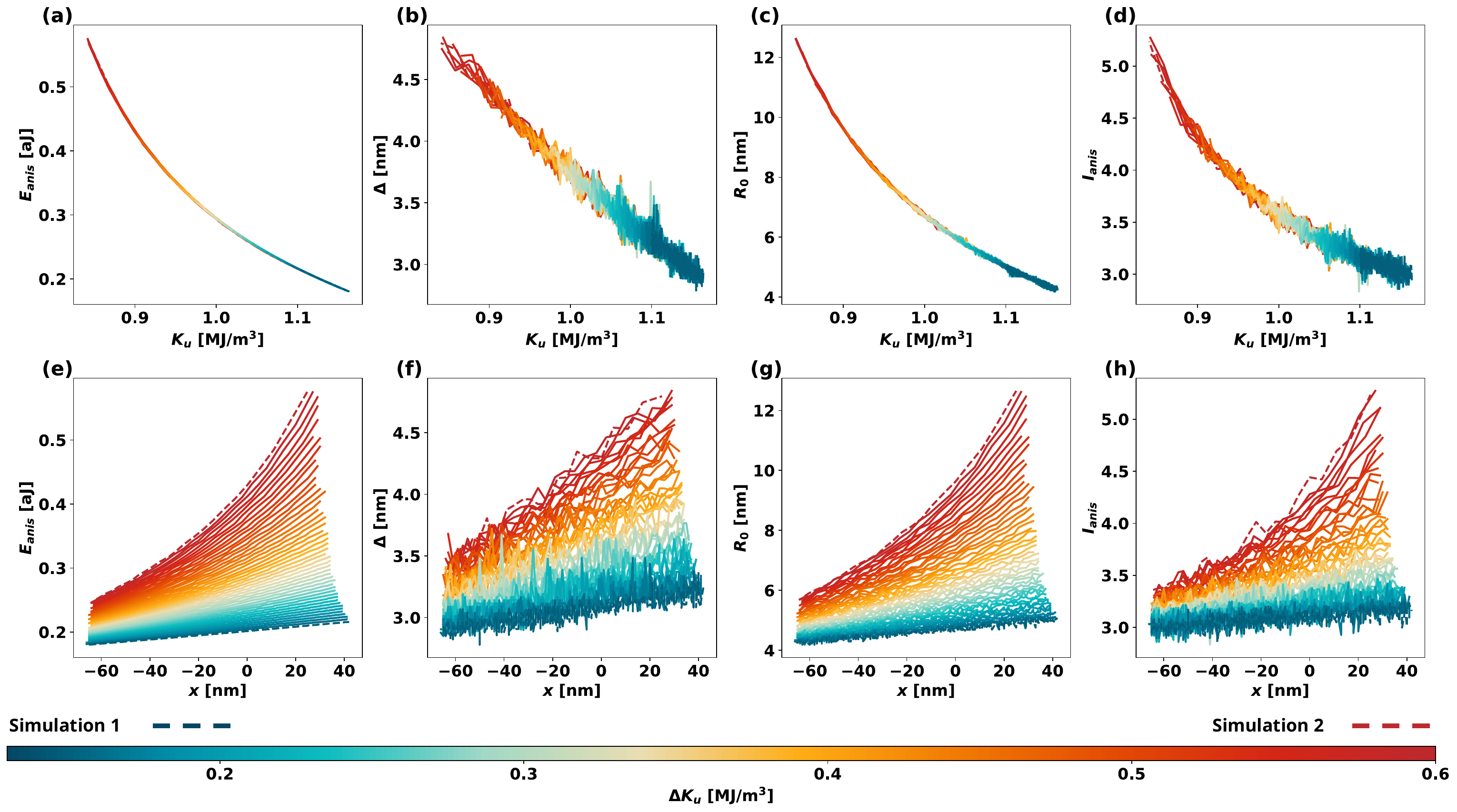} 
  \caption{Numerical analysis of skyrmion motion and deformation. The colors of the individual curves  depict the gradient strength as indicated at the bottom. The weak-gradient limit $\Delta K_\mathrm{u} = 0.15\,\mathrm{MJ}/\mathrm{m}^3$ (blue dashed line) corresponds to simulation 1 presented in Fig. \ref{fig1:FM_skyrmion}(b). The strong-gradient limit $\Delta K_\mathrm{u} = 0.60\,\mathrm{MJ}/\mathrm{m}^3$ (red dashed line) corresponds to simulation 2 presented in Fig.~\ref{fig1:FM_skyrmion}(c). The individual panels show (a) the anisotropy Energy $E_\mathrm{anis}$, (b) the domain wall width $\Delta$, (c) the skyrmion radius $R_0$ and (d) the function $I_\mathrm{anis}$ versus the perpendicular magnetic anisotropy $K_\mathrm{u}$. The images below (e-h) show the same quantities versus the $x$ coordinate of the skyrmion's center position.}
  \label{fig:quantities}
\end{figure*}

The results of the fitting procedure of $R_0(K_\mathrm{u})$ and $\Delta(K_\mathrm{u})$ as well as $I_\mathrm{anis}(K_\mathrm{u})$ and $E_\mathrm{anis}(K_\mathrm{u})$ are presented in Fig.~\ref{fig:quantities}(a-d). The respective dependencies on $x$ are shown below in panels (e-h). Particularly interesting is the strong-gradient limit (red colors), where the profile of the skyrmion changes significantly. $\Delta$ changes approximately linearly with $K_\mathrm{u}$, while $R_0$ has a non-linear form. In the investigated limits, $R_0$ changes by $ \approx 200\%$, while $\Delta$ changes by $\approx 74\%$.

In Fig.~\ref{fig:quantities}(a,e), we plot $E_\mathrm{anis}$  obtained from micromagnetic simulations, with respect to $K_\mathrm{u}$ and $x$, respectively. Upon increasing the PMA gradient, the energy landscape transitions from a linear function to a quadratic and even higher-order polynomial, despite the linear variation of $K_\mathrm{u}$. Eq.~\eqref{eq:Eanis} allows us to understand this behavior qualitatively if we expand $I_\mathrm{anis}(R_0/\Delta)\approx R_0/\Delta$ in the relevant parameter range
\begin{align}
\label{eq:expandedEanis}
E_\mathrm{anis}(x) \propto K_\mathrm{u} \cdot \Delta \cdot R_0,
\end{align}
and use the results of the fit: The domain wall width $\Delta$ depends linearly on $K_\mathrm{u}$ and the skyrmion radius $R_0$ depends quadratically on $K_\mathrm{u}$ once we approach stronger PMA gradients (red curves). This results in a fourth-order polynomial for $E_\mathrm{anis}(K_\mathrm{u})$ and $E_\mathrm{anis}(x)$. Therefore, the force $-\nabla E_\mathrm{anis}$ is not constant and results in an acceleration of the skyrmion motion.

\subsubsection{Skyrmion acceleration}

As we have established, the transition to the high-gradient regime results in a non-linear form of $E_\mathrm{anis}$ in $K_\mathrm{u}$ and $x$. However, all other energy terms are affected when $R_0$ and $\Delta$ change as well, because $E_\mathrm{ex}$ and $E_\mathrm{DMI}$ also depend on $\theta(r)$. Evaluating all the energy terms can be quite cumbersome, and some terms do not possess an obvious analytical solution~\cite{buttner2018theory}. Still, from the $K_\mathrm{u}$ dependence of $E_\mathrm{anis}$, we know that the total energy will be a polynomial with higher than linear order.

\begin{figure*}[t]

    \centering
    \includegraphics[width=\textwidth]{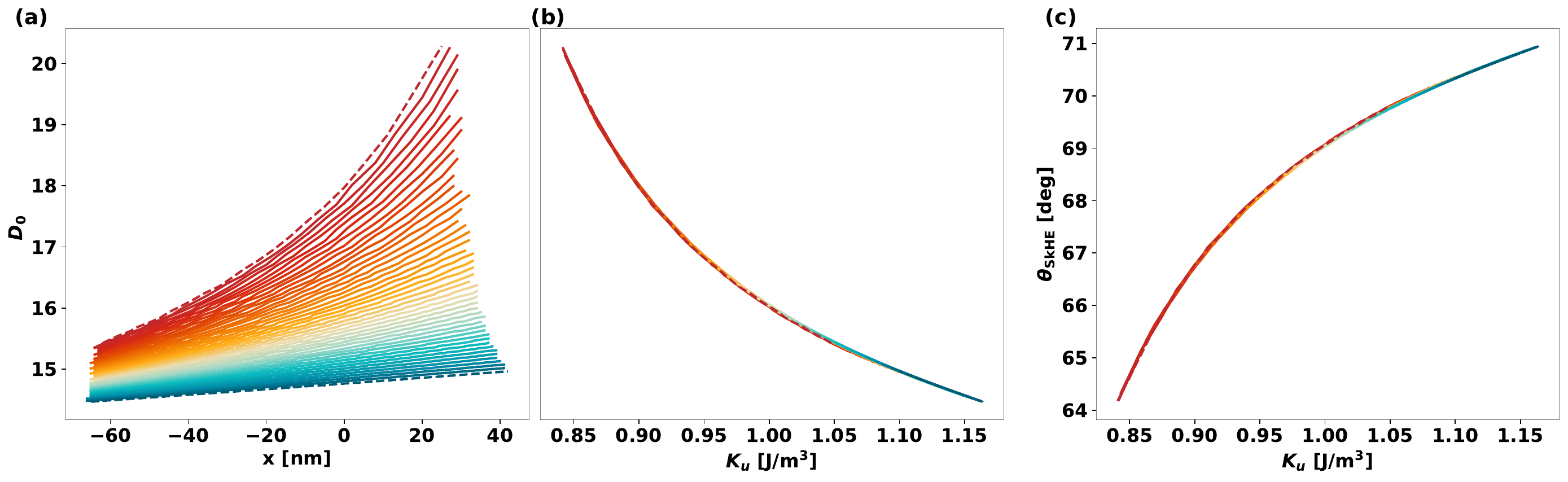}
    \caption{Skyrmion deformations causing curved trajectories. (a) Dissipative tensor $D_0$ versus the skyrmion's center's $x$ coordinate. The color corresponds to the gradient, as introduced in Fig.~\ref{fig:quantities}, and the dashed lines correspond to simulations 1 and 2 from Fig.~\ref{fig1:FM_skyrmion}(b,c). (b) Dissipative tensor $D_0$ as a function of $K_\mathrm{u}$ at the skyrmion's center position. (c) Skyrmion Hall angle as a function of $K_\mathrm{u}$ at the skyrmion's center position.}
    \label{fig:dxx}
\end{figure*}

Even though we know the dependence is of fourth or higher order in general, in our relevant parameter range, we were able to reasonably fit the total energy $E(x)$ already with a quadratic function 
\begin{align}
E(x) = A_2 x^2 +A_1 x + A_0,  \label{eq:quadratic}
\end{align}
where $A_2$, $A_1$, $A_0$ are coefficients obtained by fitting the total energy $E(x)$ vs $x$. A more detailed analysis is presented in the appendix.
This gives rise to a simple force term in the Thiele equation
\begin{align}
    -\nabla E(x) = (2A_2 x +A_1)\bm{e}_x,
\end{align}
which allows us to immediately understand why the skyrmion is accelerated.

From the Thiele equation, we arrive at
\begin{align}
    v_x(t) &= - \frac{D_0 \alpha [A_1 + 2 A_2 x(t)]}{b [(4\pi N_\mathrm{Sk})^2 + D_0^2 \alpha^2]},\notag
    \\
    v_y(t) & = \frac{4\pi N_\mathrm{Sk} [A_1 + 2 A_2 x(t)]}{b [(4\pi N_\mathrm{Sk})^2 + D_0^2 \alpha^2]}.
    \label{eq:velocities_full}
\end{align}
Calculating the time derivative yields the finite acceleration
\begin{align}
    a_x(t) &= - 2 \frac{D_0 \alpha  A_2}{b [(4\pi N_\mathrm{Sk})^2 + D_0^2 \alpha^2]}v_x(t),\notag
    \\
    a_y(t) &= 2 \frac{4\pi N_\mathrm{Sk} A_2}{b [(4\pi N_\mathrm{Sk})^2 + D_0^2 \alpha^2]}v_x(t). 
    \label{eq:accelaration}
\end{align}
As discussed, the energy will have even higher-order terms in practice. Additionally, even the dissipative tensor element $D_0$ varies with $K_\mathrm{u}$, further complicating the discussion when the PMA gradient is very strong.

Still, most importantly, we have shown that a linearly varying anisotropy can introduce an acceleration term in the Thiele equation. The skyrmion will speed up due to the deformation in the changed magnetic parameters. If the gradient is weak, the profile of the skyrmion remains roughly constant, and the skyrmion moves at a constant velocity. If the gradient increases, the skyrmion exhibits a more significant profile change from the initial to the last time step. This induces a non-linear energy landscape in $K_\mathrm{u}$ that results in the skyrmion acceleration.

\subsection{Curved trajectories and importance of the dissipative tensor}\label{sec:dissipative}

Now that we have understood why a skyrmion is accelerated, we turn to the second observation from analyzing the skyrmion motion presented in Fig.~\ref{fig1:FM_skyrmion}(d): In the strong PMA gradient, the skyrmion trajectory is curved, meaning the skyrmion Hall angle changes during propagation. This behavior cannot be explained using the equations we have presented so far. Even with the derived non-constant velocities, the skyrmion Hall effect is evaluated as before
\begin{align}
    \theta_\mathrm{SkHE} = \arctan\frac{4\pi N_\mathrm{Sk}}{\alpha D_{0}}.
\end{align}

Yet, the curved trajectory is also caused by deformation. Since $N_\mathrm{Sk}$ is a topological invariant, it is independent of continuous variations of the skyrmion profile. Therefore, the deformation only affects the dissipative tensor component $D_0$ that was assumed to be a constant in the previous discussion. However, since it is calculated from the skyrmion profile $\theta(r)$, it depends on $K_\mathrm{u}(x)$ as well.

In Fig.~\ref{fig:dxx}(a,b), we present the numerically calculated $D_0 = D_{xx} = \int ( \partial_{x} \bm{m} \cdot \partial_{x} \bm{m} ) \, \mathrm{d}^2 r$ based on the results of the previously discussed 46 micromagnetic simulations. Similar to the energy, this quantity demonstrates a dependency on the ratio $ R_0/ \Delta$. For a relatively weak PMA gradient, it exhibits a linear relationship with the skyrmion center's $x$ coordinate. However, under a stronger PMA gradient, it transitions to a non-linear behavior. This change arises from modifications in the skyrmion profile.

In the case of the strongest PMA gradient, $D_0$ changes from $\approx 15.4$ to $\approx 20.3$, which results in a change of the skyrmion Hall angle from $\approx 64.2^{\circ}$ to $\approx 69.8^{\circ}$.
The skyrmion Hall angle as a function of $K_\mathrm{u}$ is presented in Fig.~\ref{fig:dxx}(c). These results highlight the connection between $D_0$ and the PMA gradient, illustrating how changes in the gradient induce non-linear effects in the dissipative tensor. Such dependencies are crucial in understanding the behavior of skyrmions and their response to varying external conditions.

As we have seen, the gradient-driven skyrmion motion is generally non-linear and may appear difficult to control. However, this type of motion still allows us to improve the relevance of skyrmions for application, for instance, by suppressing the skyrmion Hall effect, as we shall discuss next.

\subsection{Suppression of the skyrmion Hall effect}\label{sec:suppression}

\begin{figure*}[t]
    \centering
    \includegraphics[width=\textwidth]{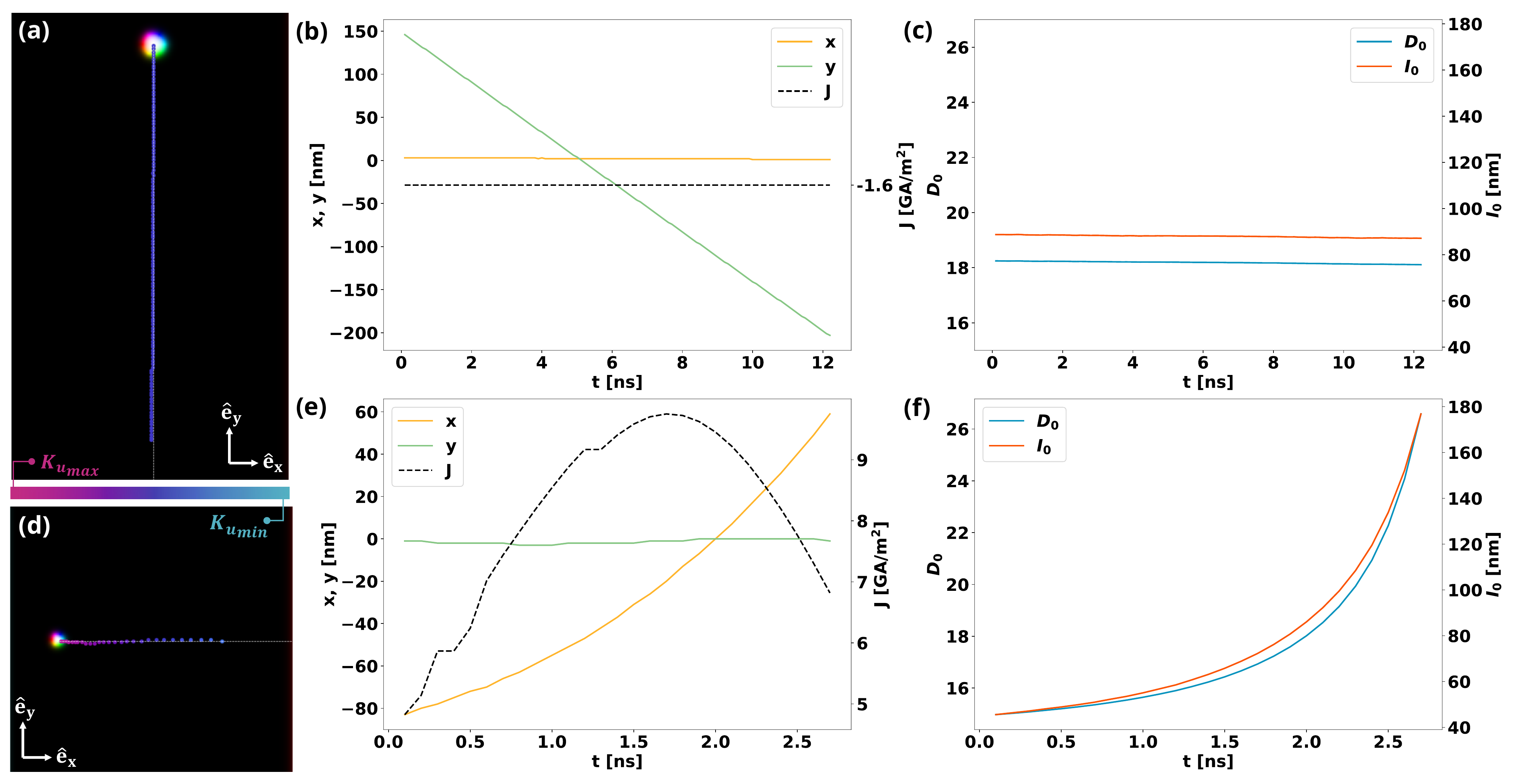}
    \caption{Motion along the current and PMA gradient direction without skyrmion Hall effect. (a-c) Motion along the current direction $-y$ (scenario 1). The PMA gradient is along $x$ and the current $j=-1.66\,\mathrm{GA}/\mathrm{m}^2$ is oriented along $-y$. (a) Trajectory based on micromagnetic simulation (purple). (b) $x$ and $y$ coordinate of the skyrmion's center position versus time. The constant velocity is $v_y = -28.4 \mathrm{m}/\mathrm{s}$. (c) Deformation of the skyrmion is measured in terms of the tensor elements $D_0$ and $I_0$ that are almost constant. (d-f) Motion along the PMA gradient direction $x$ without the skyrmion Hall effect (scenario 2). This time, the current $j$ is along $+y$ and has to be changed in magnitude due to deformations of the skyrmion in accordance with Eq.~\eqref{eq:suppresionSHEvy}. (d) Trajectory based on micromagnetic simulation (color indicates $K_\mathrm{u}$). (e) $x$ and $y$ coordinate the skyrmion's center position as well as applied current versus time. (f) Deformation of the skyrmion measured in terms of the tensor elements that change drastically from $D_0 \approx 15 $ and $I_0 \approx 45.5\,\mathrm{nm}$ to $D_0 \approx 77.5 $ and $I_0 \approx 176.8\,\mathrm{nm}$.}
    \label{fig:suppresionSHEvx}
\end{figure*}

One of the main shortcomings of foreseen skyrmion-based applications is the skyrmion Hall effect. It is caused by the topological charge of the skyrmion that enters the Thiele equation via the gyroscopic force term $4\pi N_\mathrm{Sk} \bm{e}_z\times \bm{v}$. It occurs when skyrmions are driven by SOT and, as shown in this paper, when skyrmions move in PMA gradients. In both cases, the transverse force term causes the skyrmion to not move along the respective stimulus direction but at a skyrmion Hall angle. However, if the forces related to the PMA gradient and the SOT are perpendicular to each other (or at least oriented at an angle to each other), we can achieve a suppression of the skyrmion Hall effect.

The Thiele equation, including the PMA gradient and the SOT, reads 
\begin{equation}\label{eq:Thiele_withcurrent}
    b \underline{\bm{G}} \times \bm{v} - b \underline{\bm{D}} \alpha \bm{v}   - \nabla E(x)  - B j \underline{\bm{I}} \bm{s} = 0,    
\end{equation}
where $I_{ij} = \int ( \partial_{x_i} \bm{m} \times \bm{m} )_j \,\mathrm{d}^2 r$ is the torque tensor, with $I_{xy} = -I_{yx} = I_0$ and $I_{xx} = I_{yy} = 0$ for Néel-Skyrmions. It enters the last term that accounts for the SOT. A charge current $\bm{j}$ is applied in a heavy metal layer that is interfaced with the considered FM. The spin Hall effect leads to the generation of a spin current $j_s=j\theta_\mathrm{SH}$ along $z$ with spins $\bm{s}$ oriented along $\bm{j}\times\bm{e}_z$. Therefore, by preparing the contacts for the application of the charge current along specific directions, one can control the orientation of the spins in the $xy$ plane. The constant is $ B = \hbar \Theta_\mathrm{SH} / 2 e $.


As explained, we want the forces related to the PMA gradient and the SOT to be perpendicular.
If we consider the PMA gradients along the $x$ direction, as before, we have to apply the current along $\pm y$ so that $\bm{s}$ is oriented along $x$. Eq.~\eqref{eq:Thiele_withcurrent} yields the velocities
\begin{align}
    v_x(t) &= \frac{{B I_0 j 4\pi N_\mathrm{Sk}  - D_0 A_1  \alpha - 2 A_2 D_0 \alpha x(t)}}{{b [(4\pi N_\mathrm{Sk})^2 + D_0^2 \alpha^2]}},
    \\
    v_y(t) &= \frac{{A_1 4\pi N_\mathrm{Sk}  + B D_0 I_0 j \alpha + 2 A_2 4\pi N_\mathrm{Sk} x(t)}}{{b (4\pi N_\mathrm{Sk})^2 + b D0^2 \alpha^2}}.
\end{align}

Depending on the magnitude and the sign of the applied current, the direction of motion can be tuned. Two particularly interesting scenarios correspond to a suppressed SkHE, each with distinct potential for applications: 
\begin{enumerate}
    \item The skyrmion can move along the current direction $y$, perpendicular to the anisotropy gradient.
    \item The skyrmion moves perpendicular to the current along the gradient direction $x$.
\end{enumerate}


\subsubsection{Motion along the current direction}

More relevant for an application in a racetrack storage device is scenario 1. This is because if the skyrmion propagates along the current direction and perpendicular to the gradient, the anisotropy at the skyrmion's center remains constant, and the skyrmion does not deform. Also, it means that the sample can be prepared, in principle, with an infinite length along the current direction. 

The condition for achieving this scenario is $v_x=0$ leading to
\begin{equation}
j = \frac{{D_0 \alpha (A_1 + 2 A_2 x)}}{{B I_0 4\pi N_\mathrm{Sk} }},
\label{eq:suppresionSHEvx}
\end{equation}
Figs.~\ref{fig:suppresionSHEvx}(a-c) visualize a simulation in which the skyrmion moves according to this condition. We start by writing a skyrmion at $\bm{r} = (0, 150\,\mathrm{nm})$, where $K_\mathrm{u}= 9.0\times 10^5$ J/m$^3$. The skyrmion is characterized by the tensor elements $I_0 \approx 45.5\,\mathrm{nm}$ and $D_0 \approx  18.25$. For these values, Eq.~\eqref{eq:Thiele_withcurrent} results in $j = -1.66\,\mathrm{GA}/\mathrm{m}^2$ which we have used for this simulation. This value is a constant since the $x$ coordinate of the skyrmion's center is constant in this scenario. In the simulation, the SkHE is mostly suppressed. The observed skyrmion Hall angle is only $\approx 0.3^{\circ}$. As displayed in Fig.~\ref{fig:suppresionSHEvx}(c), $D_0$ and $I_0$ change by a small amount of $\approx 0.75 \%$ and $\approx 1.8\%$, respectively. In other words, the profile remains almost unchanged due to the constrained motion along the $y$ direction, making it possible to mitigate the SkHE with a constant current. In this case, the skyrmion does not accelerate even in the strong-gradient limit. Instead, we achieve a uniform motion without the skyrmion Hall effect, ideal for racetrack applications.
 
\subsubsection{Motion along the magnetic anisotropy gradient direction}

More complicated is scenario 2, where we make the skyrmion move perfectly along the PMA gradient direction. The current is now oriented along the opposite direction as before, $+y$, to completely suppress the motion along the current direction which is perpendicular to the PMA gradient.

In Fig.~\ref{fig:suppresionSHEvx}(d-f), we present the result of a micromagnetic simulation in which the current strength was varied in time, as shown in panel (e). If this is done properly, the skyrmion moves without a skyrmion Hall effect, perfectly along the PMA gradient direction $x$. 

We started by writing a skyrmion at $\bm{r} = (0, -83\,\mathrm{nm})$, where $K_\mathrm{u}= 1.09\,\mathrm{MJ}/\mathrm{m}^3$. If a constant current is applied, the skyrmion might be able to move without a SkHE, along $x$, in the beginning. However, during the propagation, the PMA magnitude at the skyrmion's center changes causing the skyrmion to deform. This is shown in panel (f). Both $D_0$ and $I_0$ change drastically by approximately $77\%$ and $288\%$, respectively. As we have extensively discussed before, this deformation changes the skyrmion Hall angle, even in the absence of SOTs. The current has to be changed during the simulation to compensate for this variation in the propagation direction.

To achieve $v_y=0$, we can derive the current from Eq.~\eqref{eq:Thiele_withcurrent}, 
\begin{equation}
j = -\frac{{A_1 4\pi N_\mathrm{Sk} + 2 A_2 4\pi N_\mathrm{Sk} x}}{{B I_0  D_0 4\pi N_\mathrm{Sk}  \alpha}}.
\label{eq:suppresionSHEvy}
\end{equation}
At the start of the simulation, this equation yields $j = 4.82\, \mathrm{GA}/\mathrm{m}^2$. However, the current has to be changed due to the explicit $x$ dependence in the above equation and since $D_0$ and $I_0$ are also $x$ dependent. As is shown in panel (e), the current in the simulation was calculated with the time-varying quantities in Eq.~\eqref{eq:suppresionSHEvy} updated at every time step.

In summary, we have succeeded in suppressing the skyrmion Hall effect and making the skyrmion move along $x$. However, in this scenario number 2, the skyrmion deforms, which requires a changing current and leads to acceleration [panel (e)]. While a straight path along the gradient is possible, we suggest that aiming for a motion perpendicular to the gradient offers a more practical solution for controlling the skyrmion movement while successfully mitigating the SkHE. Scenario 1 circumvents the need for continuous current adjustments and allows the skyrmion to move with minimal alterations in its profile.

Still, scenario 2 might be significant for neuromorphic applications because the motion parallel to a gradient direction can mimic the excitatory process of a neuron \cite{li2017magnetic}. The skyrmion's accumulated travel distance towards a detection area in a higher PMA region mimics the increase of a biological neuron's `membrane potential'. This potential increases with a stimulus due to the accumulation of charges at the neuron's membrane. Without stimulus, these charges leak through the neuron's membrane, decreasing its membrane potential and resetting it to its initial state. By analogy, in the absence of a current, the skyrmion moves back to the region with lower PMA, losing its `accumulated traveled distance' and thereby being reset to its initial state. Suppressing the skyrmion Hall effect along the gradient could make gradient-based artificial neurons with skyrmions more reliable by avoiding skyrmion destruction at the edges while mimicking a biological neuron.

\subsection{Similar effects with different parameter gradients}\label{sec:comparison}

Before we conclude, we want to generalize our discussion. As we have shown, the skyrmion motion in an anisotropy gradient leads to a deformation of the skyrmion. This causes the skyrmion to accelerate and move along a curved trajectory. The deformation of the skyrmion is crucial for its dynamics. However, what causes this deformation is not important. Equivalent results can be achieved by considering gradients in other material parameters if they cause the skyrmion to deform and/or change size.


The above discussion can be repeated for gradients in other material properties, such as the DMI strength $D_i$ and the sample thickness. Theoretically, all of these gradients allow for an acceleration of the skyrmion motion, propagation along a curved trajectory, and potential suppression of the skyrmion Hall effect if a perpendicular current is applied, as explained before.

While extensive research on PMA gradients exists, reproducing gradients in other parameters, like DMI, could be more challenging indicated by the limited literature available. Nonetheless, generating a DMI gradient could in principle be possible, for example, by manipulating it with strain, as in Ref.~\cite{gusev2020manipulation}


\begin{figure}[t]
    \centering
    \includegraphics[width=\columnwidth]{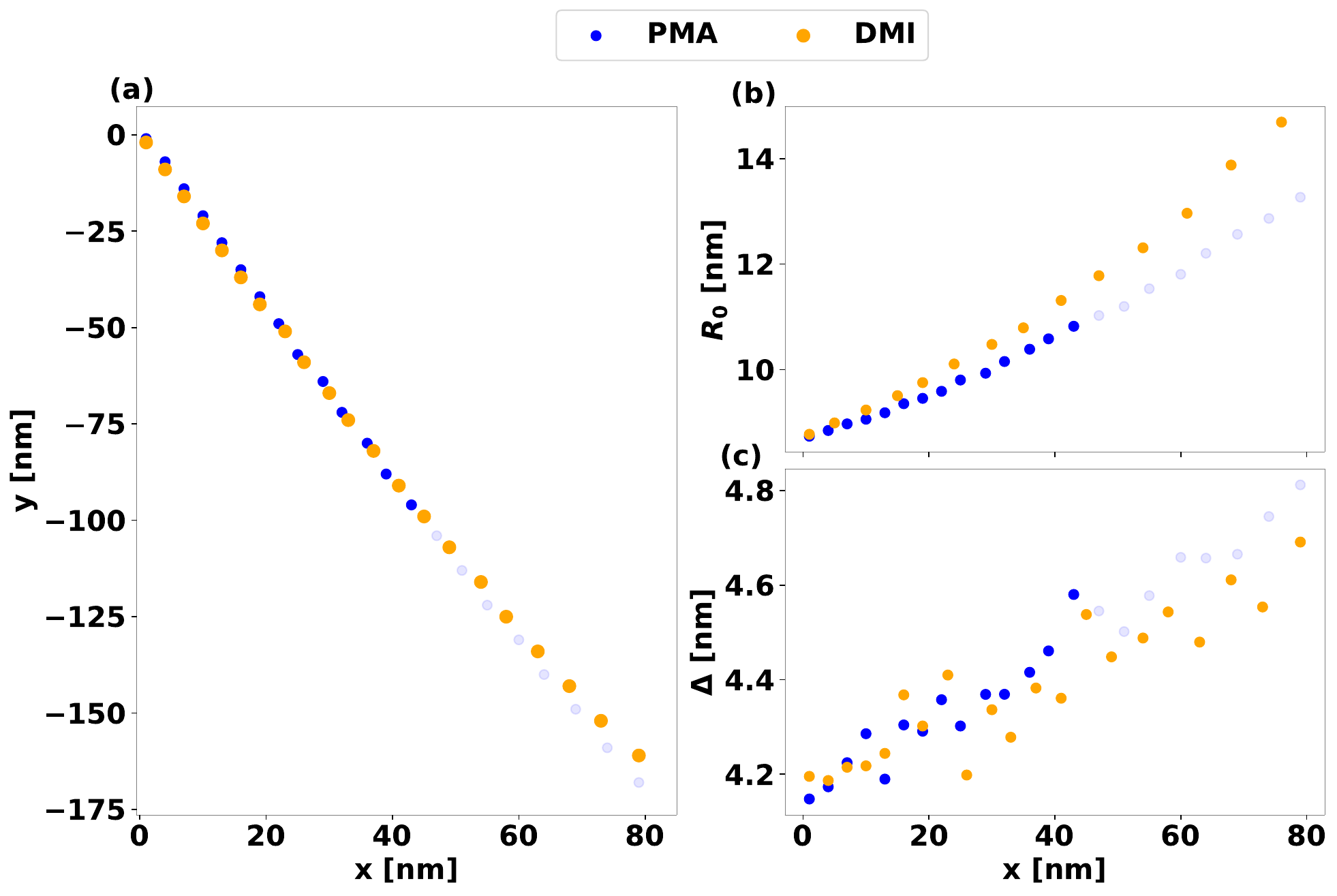}
    \caption{Comparison of the skyrmion motion in PMA (blue) and DMI gradients (orange). (a) Skyrmion trajectories. In both simulations, the skyrmion initially has the same profile as shown in (b), characterized by $R_0$ and (c) by $\Delta$.}
    \label{fig:compareDMIPMA}
\end{figure}


We could now repeat and discuss all the above calculations and simulations for a DMI gradient, but the essential point is that the DMI energy can be evaluated as
\begin{equation}
    E_\mathrm{DMI} = - 2 \pi d_z D_i \Delta I_\mathrm{DMI}(R_0/\Delta)
\end{equation}
where $I_\mathrm{DMI}$ is a non-elementary function (see supplementary material of Ref.~\onlinecite{buttner2018theory}). By analogy with our discussion of the PMA gradient before, $E_\mathrm{DMI}$ does not only depend explicitly on $D_i$ but also implicitly via the domain wall width $\Delta$ and the skyrmion radius $R_0$ that account for deformations of the skyrmion while propagating through the different regions with changing $D_i$. This means $-\nabla E_\mathrm{DMI}$ is a polynomial of higher than linear order in $D_i$ (and therefore also in $x$), giving rise to a non-constant force. Note that, as before, it is not sufficient to consider only $E_\mathrm{DMI}$, but instead, the total energy $E$ has to be considered.

Based on this finding, we can qualitatively reproduce all three key findings of this paper. (i) The skyrmion moves linearly in a weak DMI gradient and accelerates in a stronger gradient. (ii) If the DMI gradient is strong, the skyrmion Hall angle changes during the propagation giving rise to a curved trajectory. (iii) If we apply a current perpendicular to the DMI gradient that gives rise to a SOT, we can compensate the skyrmion Hall effect if the current strength is chosen appropriately. 

Skyrmions move similarly in a PMA gradient and a DMI gradient, except for one qualitative difference: The skyrmion has lower energy in the region with a higher DMI. In other words, it moves from $D_{i_\mathrm{min}}$ to $D_{i_\mathrm{max}}$, while it moved from $K_{\mathrm{u}_\mathrm{max}}$ to $K_{\mathrm{u}_\mathrm{min}}$ before. A stronger PMA favors a `stiff' collinear spin texture, while a stronger DMI favors a non-colinear spin texture. 

When comparing the motion of the skyrmion driven by PMA and DMI gradients quantitatively, it is essential to realize that the motion is not only affected by the value of the gradient $\Delta K_\mathrm{u}/l$ and $\Delta D_i/l$ but also by the starting value of the parameters $K_\mathrm{u}$ and $D_i$. While all the results presented are qualitatively the same for various materials with different magnetic parameters, the following quantitative analysis is only meaningful for our particular material with the chosen set of parameters.

To compare the motion of a skyrmion in a PMA gradient and a DMI gradient, we start with a skyrmion for which $K_\mathrm{u}=0.96$ MJ/m$^3$ and $D_i= 4.2$ mJ/m$^2$ at the center. Then we consider (i) a PMA gradient with $K_{\mathrm{u}_\mathrm{max}} = 1.12$  MJ/m$^3$ and $K_{\mathrm{u}_\mathrm{min}} = 0.8$  MJ/m$^3$ and (ii) repeat the simulation with a DMI gradient instead where $D_{\mathrm{i}_\mathrm{max}} = 3.6$  mJ/m$^2$ and $D_{\mathrm{i}_\mathrm{min}} = 3.0$  mJ/m$^2$. Both gradients correspond to a relative change of the parameter of $40\%$ throughout the sample. 

The results of the two simulations are presented in Fig.~\ref{fig:compareDMIPMA}. The skyrmion starts roughly with the same profile characterized by $R_0 \approx 8.7$ nm and $\Delta \approx 4.1$ nm. However, despite the identical initial conditions, our simulations suggest a different rate at which the profile changes. In the DMI gradient, the skyrmion reaches the edges after $8$ ns, and in the PMA gradient after $14$ ns, with different $R_0$ and $\Delta$. Thus, a DMI gradient might enable the skyrmion to achieve higher velocities in our considered material. 


In summary, a DMI gradient behaves similarly to a PMA gradient. It exhibits significant profile changes and allows for acceleration, curved trajectories, and a compensated skyrmion Hall effect. Although further experimental validations for such gradient are still needed, a DMI gradient could be more desirable for gradient-based devices that require a fast operation.

\begin{figure*}[t]
    \centering
    \includegraphics[width=0.9\textwidth]{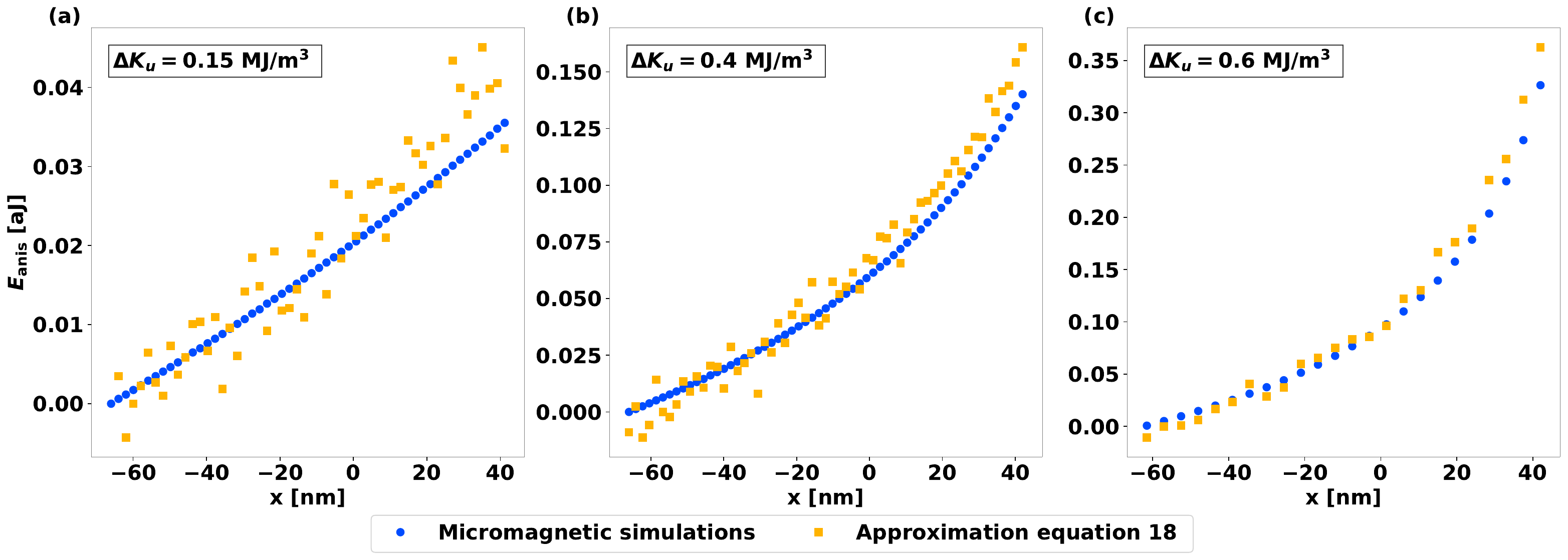}
    \caption{Approximation of the anisotropy energy. The figure illustrates the approximated anisotropy energy Eq.~\eqref{eq:Eanis} (orange squares), calculated by numerically evaluating $I_\mathrm{anis}$ based on Eq.~\eqref{eq:Ianis}. It is compared to the anisotropy energy [same as Fig.~\ref{fig:quantities}(e)] obtained by micromagnetic simulations (blue dots). The energies are compared for (a) the weak-gradient limit, (b) a transition point between the weak- and strong-gradient limits (cf. color bar in Fig.~\ref{fig:quantities}), and (c) the strong-gradient limit.}
    \label{fig:fitEanis}
\end{figure*}

\begin{figure}[t]
    \centering
    \includegraphics[width=\columnwidth]{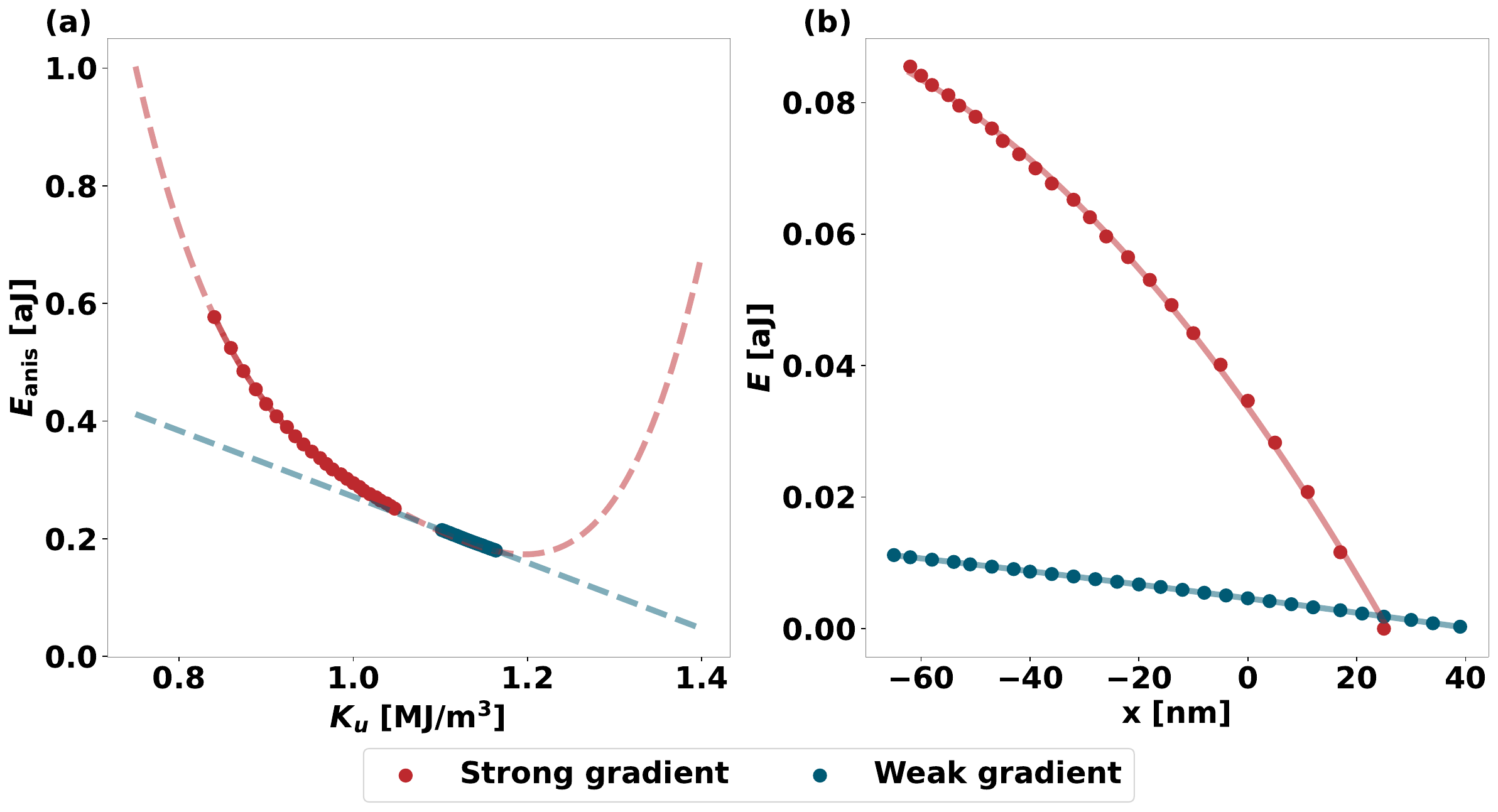}
    \caption{Dependence of the anisotropy energy and the total energy on $K_\mathrm{u}$. The figure illustrates the energy variations in the strong-gradient limit (red) and the weak-gradient limit (blue). The points (dots) represent the data obtained from micromagnetic simulations [same as Fig.~\ref{fig:quantities}(a)], while the lines correspond to polynomial fits. In (a), the anisotropy energy is fitted to a fourth-order polynomial for the strong gradient case and to a linear polynomial for the weak gradient case. The dashed curves represent the extrapolated fit for $K_\mathrm{u}$ values beyond our set of parameters. In (b), the total energy is fitted versus $x$ with a second-order polynomial for the strong gradient scenario and a linear polynomial for the weak gradient.}
    \label{fig:TotalEnergyFit}
\end{figure}

\section{Conclusions}\label{sec:conclusion}

In summary, we have investigated the motion of a magnetic skyrmion in gradients of material parameters. For the majority of our discussion, we have analyzed gradients of the perpendicular magnetic anisotropy $K_\mathrm{u}$. The skyrmion moves from a high anisotropy region to a lower one. As long as the gradient is so small that it does not deform during propagation, the driving force is constant -- like the driving force corresponding to SOTs. The skyrmion moves uniformly but at an angle with respect to the gradient direction.
Once the PMA gradient is so large that the skyrmion changes its size and shape during the propagation from one side to the other, the skyrmion (i) experiences an acceleration, as discussed in Sec.~\ref{sec:pmagradient}, and (ii) it moves along a curved trajectory, as discussed in Sec.~\ref{sec:dissipative}. 

Furthermore, when we apply a current to drive the skyrmion by SOT, we can manipulate the motion direction by tuning the current's magnitude and sign. By choosing the appropriate values, we can completely suppress the skyrmion Hall effect so that the skyrmion moves either along the current or gradient direction. As discussed at the end of Sec.~\ref{sec:suppression}, the skyrmion Hall effect could be detrimental to applications of skyrmions in technology, and suppressing it enhances the potential for spintronic and neuromorphic applications

As discussed in Sec.~\ref{sec:comparison}, the same results can be achieved using other material parameters' gradients. Noteworthy is the DMI gradient, as it may be present in a sample with varying stochiometry or a sample with interfacial DMI and a varying thickness. In this case, the effective DMI, averaged over the whole thickness, changes linearly.
In practice, it is hardly possible that only a single material parameter gradient is present. Instead, multiple magnetic parameters will vary throughout the sample, and the effect of several parameter gradients will act simultaneously.

Our results are also significant from a fundamental point of view. In Ref.~\cite{schutte2014inertia, buttner2015dynamics}, an effective skyrmion mass has been defined via a generalized Thiele equation. In our simulations, the deformation of a skyrmion leads to acceleration, which might also suggest the presence of a skyrmion mass and inertia. However, it is worth noting that the Thiele equation~\eqref{eq:Thiele} used in this work is a differential equation of first-order in time. Therefore, it does not account for mass and we observe no inertia in our micromagnetic simulations. Still, acceleration is allowed as soon as the gradient of the total energy is not a constant.


\appendix

\section{Details about the anisotropy energy}

Eq.~\eqref{eq:Eanis} relates the anisotropy energy to the skyrmion radius $R_0$ and the domain wall width $\Delta$ as derived in Ref.~\onlinecite{buttner2018theory}. It contains the function
\begin{widetext}
\begin{equation}
\label{eq:Ianis}
    I_\mathrm{anis}(R_0/\Delta) = \frac{1}{2} \frac{\cosh{(R_0/\Delta)}}{[\sinh{(R_0/\Delta)}]^3} \Bigg[ - \mathrm{Li}_2\big(- e^{-2(R_0/\Delta)} \big) + \mathrm{Li}_2\big(- e^{2(R_0/\Delta)} \big) + 2 \ln{\big( 2 \cosh{(R_0/\Delta)}\big)} \sinh{(2 R_0/\Delta)} \Bigg],
\end{equation}
\end{widetext}
where $\mathrm{Li}_2$ is the dilogarithm.

Fig.~\ref{fig:fitEanis} shows that the anisotropy energy $E_\mathrm{anis}(x)$ can indeed be approximated by Eq.~\eqref{eq:Eanis}.
The micromagnetic data (blue) is compared with the result of Eq.~\eqref{eq:Eanis} (orange) evaluated numerically with $R_0$ and $\Delta$ obtained by fitting the micromagnetic skyrmion profiles (cf. Fig.~\ref{fig:quantities}). Since we are interested in the energy change, the plots are adjusted by setting the minimum $E_\mathrm{anis}^{0}$ as the reference.

It is worth noting that evaluating $\Delta$ from the data can be challenging for the weak-gradient limit due to the skyrmions' small sizes and deformations. This results in noise and less accurate fitting. However, a better fit is achieved as $\Delta$ increases, such as in panels (b) and (c).

\section{Total Energy}

In the previous appendix, we verified that the anisotropy energy can be approximated by Eq.~\eqref{eq:Eanis}. As explained in the main text, we can fit $R_0$ and $\Delta$ with respect to $K_\mathrm{u}$ and find that the anisotropy energy behaves like a fourth-order polynomial in $K_\mathrm{u}$ [cf. Eq.~\eqref{eq:expandedEanis}].


Fig.~\ref{fig:TotalEnergyFit}(a) shows $E_\mathrm{anis}$ as a function of $K_\mathrm{u}$. A fourth-order polynomial fit is applied for the strong-gradient scenario (red). The fit reproduces the data very well; however, in the considered parameter range, even a quadratic function can capture the non-linear behavior well enough. For the weak-gradient case (blue), the data is even linear, which is why we have fitted a linear function in this case. Note that the blue data points agree well with the fourth-order polynomial (red) because it behaves linearly in this particular range of parameters.



Since $-\nabla E$ enters the equation of motion, we have to analyze the total energy and not just the anisotropy energy. Fig.~\ref{fig:TotalEnergyFit}(b) shows that it is sufficient to fit the energy by a quadratic function in $K_\mathrm{u}$ or $x$ in the strong-gradient case (red) and by a linear function in the weak-gradient case (blue). This is why we have limited the analysis of the skyrmion acceleration to quadratic-order energy terms [cf. Eq.~\eqref{eq:quadratic}]. This allows us to provide a simple description of the force term for the Thiele equation without a lack of accuracy. The fit yields the coefficients  $A_2 \approx -5.03 \times 10^{-6}$ J/m$^2$ and $A_1 \approx -1.13 \times 10^{-12}$ J/m that were used in Sec.~\ref{sec:suppression}. In the weak-gradient case, we find that a first-order polynomial is a good fit yielding only the coefficient $A_1 \approx -1.07 \times 10^{-13}$ J/m.

\begin{acknowledgments}
This work is supported by SFB TRR 227 of Deutsche Forschungsgemeinschaft (DFG) and SPEAR ITN.
This project has received funding from the European Union’s Horizon 2020 research and innovation program under the Marie Skłodowska-Curie grant agreement No 955671.
\end{acknowledgments}

\bibliography{references.bib} 

\begin{thebibliography}{29}%
\makeatletter
\providecommand \@ifxundefined [1]{%
 \@ifx{#1\undefined}
}%
\providecommand \@ifnum [1]{%
 \ifnum #1\expandafter \@firstoftwo
 \else \expandafter \@secondoftwo
 \fi
}%
\providecommand \@ifx [1]{%
 \ifx #1\expandafter \@firstoftwo
 \else \expandafter \@secondoftwo
 \fi
}%
\providecommand \natexlab [1]{#1}%
\providecommand \enquote  [1]{``#1''}%
\providecommand \bibnamefont  [1]{#1}%
\providecommand \bibfnamefont [1]{#1}%
\providecommand \citenamefont [1]{#1}%
\providecommand \href@noop [0]{\@secondoftwo}%
\providecommand \href [0]{\begingroup \@sanitize@url \@href}%
\providecommand \@href[1]{\@@startlink{#1}\@@href}%
\providecommand \@@href[1]{\endgroup#1\@@endlink}%
\providecommand \@sanitize@url [0]{\catcode `\\12\catcode `\$12\catcode
  `\&12\catcode `\#12\catcode `\^12\catcode `\_12\catcode `\%12\relax}%
\providecommand \@@startlink[1]{}%
\providecommand \@@endlink[0]{}%
\providecommand \url  [0]{\begingroup\@sanitize@url \@url }%
\providecommand \@url [1]{\endgroup\@href {#1}{\urlprefix }}%
\providecommand \urlprefix  [0]{URL }%
\providecommand \Eprint [0]{\href }%
\providecommand \doibase [0]{https://doi.org/}%
\providecommand \selectlanguage [0]{\@gobble}%
\providecommand \bibinfo  [0]{\@secondoftwo}%
\providecommand \bibfield  [0]{\@secondoftwo}%
\providecommand \translation [1]{[#1]}%
\providecommand \BibitemOpen [0]{}%
\providecommand \bibitemStop [0]{}%
\providecommand \bibitemNoStop [0]{.\EOS\space}%
\providecommand \EOS [0]{\spacefactor3000\relax}%
\providecommand \BibitemShut  [1]{\csname bibitem#1\endcsname}%
\let\auto@bib@innerbib\@empty
\bibitem [{\citenamefont {Bogdanov}\ and\ \citenamefont
  {Yablonskii}(1989)}]{bogdanov1989thermodynamically}%
  \BibitemOpen
  \bibfield  {author} {\bibinfo {author} {\bibfnamefont {A.~N.}\ \bibnamefont
  {Bogdanov}}\ and\ \bibinfo {author} {\bibfnamefont {D.}~\bibnamefont
  {Yablonskii}},\ }\bibfield  {title} {\bibinfo {title} {Thermodynamically
  stable “vortices” in magnetically ordered crystals. {T}he mixed state of
  magnets.},\ }\href@noop {} {\bibfield  {journal} {\bibinfo  {journal} {Zh.
  Eksp. Teor. Fiz}\ }\textbf {\bibinfo {volume} {95}},\ \bibinfo {pages} {178}
  (\bibinfo {year} {1989})}\BibitemShut {NoStop}%
\bibitem [{\citenamefont {Nagaosa}\ and\ \citenamefont
  {Tokura}(2013)}]{nagaosa2013topological}%
  \BibitemOpen
  \bibfield  {author} {\bibinfo {author} {\bibfnamefont {N.}~\bibnamefont
  {Nagaosa}}\ and\ \bibinfo {author} {\bibfnamefont {Y.}~\bibnamefont
  {Tokura}},\ }\bibfield  {title} {\bibinfo {title} {Topological properties and
  dynamics of magnetic skyrmions},\ }\href@noop {} {\bibfield  {journal}
  {\bibinfo  {journal} {Nature Nanotechnology}\ }\textbf {\bibinfo {volume}
  {8}},\ \bibinfo {pages} {899} (\bibinfo {year} {2013})}\BibitemShut {NoStop}%
\bibitem [{\citenamefont {Fert}\ \emph {et~al.}(2013)\citenamefont {Fert},
  \citenamefont {Cros},\ and\ \citenamefont {Sampaio}}]{fert2013skyrmions}%
  \BibitemOpen
  \bibfield  {author} {\bibinfo {author} {\bibfnamefont {A.}~\bibnamefont
  {Fert}}, \bibinfo {author} {\bibfnamefont {V.}~\bibnamefont {Cros}},\ and\
  \bibinfo {author} {\bibfnamefont {J.}~\bibnamefont {Sampaio}},\ }\bibfield
  {title} {\bibinfo {title} {Skyrmions on the track},\ }\href@noop {}
  {\bibfield  {journal} {\bibinfo  {journal} {Nature Nanotechnology}\ }\textbf
  {\bibinfo {volume} {8}},\ \bibinfo {pages} {152} (\bibinfo {year}
  {2013})}\BibitemShut {NoStop}%
\bibitem [{\citenamefont {Li}\ \emph {et~al.}(2021)\citenamefont {Li},
  \citenamefont {Kang}, \citenamefont {Zhang}, \citenamefont {Nie},
  \citenamefont {Zhou}, \citenamefont {Wang},\ and\ \citenamefont
  {Zhao}}]{li2021magnetic}%
  \BibitemOpen
  \bibfield  {author} {\bibinfo {author} {\bibfnamefont {S.}~\bibnamefont
  {Li}}, \bibinfo {author} {\bibfnamefont {W.}~\bibnamefont {Kang}}, \bibinfo
  {author} {\bibfnamefont {X.}~\bibnamefont {Zhang}}, \bibinfo {author}
  {\bibfnamefont {T.}~\bibnamefont {Nie}}, \bibinfo {author} {\bibfnamefont
  {Y.}~\bibnamefont {Zhou}}, \bibinfo {author} {\bibfnamefont {K.~L.}\
  \bibnamefont {Wang}},\ and\ \bibinfo {author} {\bibfnamefont
  {W.}~\bibnamefont {Zhao}},\ }\bibfield  {title} {\bibinfo {title} {Magnetic
  skyrmions for unconventional computing},\ }\href@noop {} {\bibfield
  {journal} {\bibinfo  {journal} {Materials Horizons}\ }\textbf {\bibinfo
  {volume} {8}},\ \bibinfo {pages} {854} (\bibinfo {year} {2021})}\BibitemShut
  {NoStop}%
\bibitem [{\citenamefont {Li}\ \emph {et~al.}(2017)\citenamefont {Li},
  \citenamefont {Kang}, \citenamefont {Huang}, \citenamefont {Zhang},
  \citenamefont {Zhou},\ and\ \citenamefont {Zhao}}]{li2017magnetic}%
  \BibitemOpen
  \bibfield  {author} {\bibinfo {author} {\bibfnamefont {S.}~\bibnamefont
  {Li}}, \bibinfo {author} {\bibfnamefont {W.}~\bibnamefont {Kang}}, \bibinfo
  {author} {\bibfnamefont {Y.}~\bibnamefont {Huang}}, \bibinfo {author}
  {\bibfnamefont {X.}~\bibnamefont {Zhang}}, \bibinfo {author} {\bibfnamefont
  {Y.}~\bibnamefont {Zhou}},\ and\ \bibinfo {author} {\bibfnamefont
  {W.}~\bibnamefont {Zhao}},\ }\bibfield  {title} {\bibinfo {title} {Magnetic
  skyrmion-based artificial neuron device},\ }\href@noop {} {\bibfield
  {journal} {\bibinfo  {journal} {Nanotechnology}\ }\textbf {\bibinfo {volume}
  {28}},\ \bibinfo {pages} {31LT01} (\bibinfo {year} {2017})}\BibitemShut
  {NoStop}%
\bibitem [{\citenamefont {de~Assis}\ \emph {et~al.}(2023)\citenamefont
  {de~Assis}, \citenamefont {Mertig},\ and\ \citenamefont
  {G{\"o}bel}}]{de2023biskyrmion}%
  \BibitemOpen
  \bibfield  {author} {\bibinfo {author} {\bibfnamefont {I.~R.}\ \bibnamefont
  {de~Assis}}, \bibinfo {author} {\bibfnamefont {I.}~\bibnamefont {Mertig}},\
  and\ \bibinfo {author} {\bibfnamefont {B.}~\bibnamefont {G{\"o}bel}},\
  }\bibfield  {title} {\bibinfo {title} {Biskyrmion-based artificial neuron},\
  }\href@noop {} {\bibfield  {journal} {\bibinfo  {journal} {Neuromorphic
  Computing and Engineering}\ }\textbf {\bibinfo {volume} {3}},\ \bibinfo
  {pages} {014012} (\bibinfo {year} {2023})}\BibitemShut {NoStop}%
\bibitem [{\citenamefont {G{\"o}bel}\ \emph {et~al.}(2021)\citenamefont
  {G{\"o}bel}, \citenamefont {Mertig},\ and\ \citenamefont
  {Tretiakov}}]{gobel2021beyond}%
  \BibitemOpen
  \bibfield  {author} {\bibinfo {author} {\bibfnamefont {B.}~\bibnamefont
  {G{\"o}bel}}, \bibinfo {author} {\bibfnamefont {I.}~\bibnamefont {Mertig}},\
  and\ \bibinfo {author} {\bibfnamefont {O.~A.}\ \bibnamefont {Tretiakov}},\
  }\bibfield  {title} {\bibinfo {title} {Beyond skyrmions: Review and
  perspectives of alternative magnetic quasiparticles},\ }\href@noop {}
  {\bibfield  {journal} {\bibinfo  {journal} {Physics Reports}\ }\textbf
  {\bibinfo {volume} {895}},\ \bibinfo {pages} {1} (\bibinfo {year}
  {2021})}\BibitemShut {NoStop}%
\bibitem [{\citenamefont {Heinze}\ \emph {et~al.}(2011)\citenamefont {Heinze},
  \citenamefont {Von~Bergmann}, \citenamefont {Menzel}, \citenamefont {Brede},
  \citenamefont {Kubetzka}, \citenamefont {Wiesendanger}, \citenamefont
  {Bihlmayer},\ and\ \citenamefont {Bl{\"u}gel}}]{heinze2011spontaneous}%
  \BibitemOpen
  \bibfield  {author} {\bibinfo {author} {\bibfnamefont {S.}~\bibnamefont
  {Heinze}}, \bibinfo {author} {\bibfnamefont {K.}~\bibnamefont
  {Von~Bergmann}}, \bibinfo {author} {\bibfnamefont {M.}~\bibnamefont
  {Menzel}}, \bibinfo {author} {\bibfnamefont {J.}~\bibnamefont {Brede}},
  \bibinfo {author} {\bibfnamefont {A.}~\bibnamefont {Kubetzka}}, \bibinfo
  {author} {\bibfnamefont {R.}~\bibnamefont {Wiesendanger}}, \bibinfo {author}
  {\bibfnamefont {G.}~\bibnamefont {Bihlmayer}},\ and\ \bibinfo {author}
  {\bibfnamefont {S.}~\bibnamefont {Bl{\"u}gel}},\ }\bibfield  {title}
  {\bibinfo {title} {Spontaneous atomic-scale magnetic skyrmion lattice in two
  dimensions},\ }\href@noop {} {\bibfield  {journal} {\bibinfo  {journal}
  {Nature Physics}\ }\textbf {\bibinfo {volume} {7}},\ \bibinfo {pages} {713}
  (\bibinfo {year} {2011})}\BibitemShut {NoStop}%
\bibitem [{\citenamefont {M{\"u}hlbauer}\ \emph {et~al.}(2009)\citenamefont
  {M{\"u}hlbauer}, \citenamefont {Binz}, \citenamefont {Jonietz}, \citenamefont
  {Pfleiderer}, \citenamefont {Rosch}, \citenamefont {Neubauer}, \citenamefont
  {Georgii},\ and\ \citenamefont {B{\"o}ni}}]{muhlbauer2009skyrmion}%
  \BibitemOpen
  \bibfield  {author} {\bibinfo {author} {\bibfnamefont {S.}~\bibnamefont
  {M{\"u}hlbauer}}, \bibinfo {author} {\bibfnamefont {B.}~\bibnamefont {Binz}},
  \bibinfo {author} {\bibfnamefont {F.}~\bibnamefont {Jonietz}}, \bibinfo
  {author} {\bibfnamefont {C.}~\bibnamefont {Pfleiderer}}, \bibinfo {author}
  {\bibfnamefont {A.}~\bibnamefont {Rosch}}, \bibinfo {author} {\bibfnamefont
  {A.}~\bibnamefont {Neubauer}}, \bibinfo {author} {\bibfnamefont
  {R.}~\bibnamefont {Georgii}},\ and\ \bibinfo {author} {\bibfnamefont
  {P.}~\bibnamefont {B{\"o}ni}},\ }\bibfield  {title} {\bibinfo {title}
  {Skyrmion lattice in a chiral magnet},\ }\href@noop {} {\bibfield  {journal}
  {\bibinfo  {journal} {Science}\ }\textbf {\bibinfo {volume} {323}},\ \bibinfo
  {pages} {915} (\bibinfo {year} {2009})}\BibitemShut {NoStop}%
\bibitem [{\citenamefont {Nayak}\ \emph {et~al.}(2017)\citenamefont {Nayak},
  \citenamefont {Kumar}, \citenamefont {Ma}, \citenamefont {Werner},
  \citenamefont {Pippel}, \citenamefont {Sahoo}, \citenamefont {Damay},
  \citenamefont {R{\"o}{\ss}ler}, \citenamefont {Felser},\ and\ \citenamefont
  {Parkin}}]{nayak2017magnetic}%
  \BibitemOpen
  \bibfield  {author} {\bibinfo {author} {\bibfnamefont {A.~K.}\ \bibnamefont
  {Nayak}}, \bibinfo {author} {\bibfnamefont {V.}~\bibnamefont {Kumar}},
  \bibinfo {author} {\bibfnamefont {T.}~\bibnamefont {Ma}}, \bibinfo {author}
  {\bibfnamefont {P.}~\bibnamefont {Werner}}, \bibinfo {author} {\bibfnamefont
  {E.}~\bibnamefont {Pippel}}, \bibinfo {author} {\bibfnamefont
  {R.}~\bibnamefont {Sahoo}}, \bibinfo {author} {\bibfnamefont
  {F.}~\bibnamefont {Damay}}, \bibinfo {author} {\bibfnamefont {U.~K.}\
  \bibnamefont {R{\"o}{\ss}ler}}, \bibinfo {author} {\bibfnamefont
  {C.}~\bibnamefont {Felser}},\ and\ \bibinfo {author} {\bibfnamefont {S.~S.}\
  \bibnamefont {Parkin}},\ }\bibfield  {title} {\bibinfo {title} {Magnetic
  antiskyrmions above room temperature in tetragonal {H}eusler materials},\
  }\href@noop {} {\bibfield  {journal} {\bibinfo  {journal} {Nature}\ }\textbf
  {\bibinfo {volume} {548}},\ \bibinfo {pages} {561} (\bibinfo {year}
  {2017})}\BibitemShut {NoStop}%
\bibitem [{\citenamefont {Jiang}\ \emph {et~al.}(2017)\citenamefont {Jiang},
  \citenamefont {Zhang}, \citenamefont {Yu}, \citenamefont {Zhang},
  \citenamefont {Wang}, \citenamefont {Benjamin~Jungfleisch}, \citenamefont
  {Pearson}, \citenamefont {Cheng}, \citenamefont {Heinonen}, \citenamefont
  {Wang} \emph {et~al.}}]{jiang2017direct0}%
  \BibitemOpen
  \bibfield  {author} {\bibinfo {author} {\bibfnamefont {W.}~\bibnamefont
  {Jiang}}, \bibinfo {author} {\bibfnamefont {X.}~\bibnamefont {Zhang}},
  \bibinfo {author} {\bibfnamefont {G.}~\bibnamefont {Yu}}, \bibinfo {author}
  {\bibfnamefont {W.}~\bibnamefont {Zhang}}, \bibinfo {author} {\bibfnamefont
  {X.}~\bibnamefont {Wang}}, \bibinfo {author} {\bibfnamefont {M.}~\bibnamefont
  {Benjamin~Jungfleisch}}, \bibinfo {author} {\bibfnamefont {J.~E.}\
  \bibnamefont {Pearson}}, \bibinfo {author} {\bibfnamefont {X.}~\bibnamefont
  {Cheng}}, \bibinfo {author} {\bibfnamefont {O.}~\bibnamefont {Heinonen}},
  \bibinfo {author} {\bibfnamefont {K.~L.}\ \bibnamefont {Wang}}, \emph
  {et~al.},\ }\bibfield  {title} {\bibinfo {title} {Direct observation of the
  skyrmion {H}all effect},\ }\href@noop {} {\bibfield  {journal} {\bibinfo
  {journal} {Nature Physics}\ }\textbf {\bibinfo {volume} {13}},\ \bibinfo
  {pages} {162} (\bibinfo {year} {2017})}\BibitemShut {NoStop}%
\bibitem [{\citenamefont {Litzius}\ \emph {et~al.}(2017)\citenamefont
  {Litzius}, \citenamefont {Lemesh}, \citenamefont {Kr{\"u}ger}, \citenamefont
  {Bassirian}, \citenamefont {Caretta}, \citenamefont {Richter}, \citenamefont
  {B{\"u}ttner}, \citenamefont {Sato}, \citenamefont {Tretiakov}, \citenamefont
  {F{\"o}rster} \emph {et~al.}}]{litzius2017skyrmion}%
  \BibitemOpen
  \bibfield  {author} {\bibinfo {author} {\bibfnamefont {K.}~\bibnamefont
  {Litzius}}, \bibinfo {author} {\bibfnamefont {I.}~\bibnamefont {Lemesh}},
  \bibinfo {author} {\bibfnamefont {B.}~\bibnamefont {Kr{\"u}ger}}, \bibinfo
  {author} {\bibfnamefont {P.}~\bibnamefont {Bassirian}}, \bibinfo {author}
  {\bibfnamefont {L.}~\bibnamefont {Caretta}}, \bibinfo {author} {\bibfnamefont
  {K.}~\bibnamefont {Richter}}, \bibinfo {author} {\bibfnamefont
  {F.}~\bibnamefont {B{\"u}ttner}}, \bibinfo {author} {\bibfnamefont
  {K.}~\bibnamefont {Sato}}, \bibinfo {author} {\bibfnamefont {O.~A.}\
  \bibnamefont {Tretiakov}}, \bibinfo {author} {\bibfnamefont {J.}~\bibnamefont
  {F{\"o}rster}}, \emph {et~al.},\ }\bibfield  {title} {\bibinfo {title}
  {Skyrmion {H}all effect revealed by direct time-resolved {X}-ray
  microscopy},\ }\href@noop {} {\bibfield  {journal} {\bibinfo  {journal}
  {Nature Physics}\ }\textbf {\bibinfo {volume} {13}},\ \bibinfo {pages} {170}
  (\bibinfo {year} {2017})}\BibitemShut {NoStop}%
\bibitem [{\citenamefont {Juge}\ \emph {et~al.}(2021)\citenamefont {Juge},
  \citenamefont {Bairagi}, \citenamefont {Rana}, \citenamefont {Vogel},
  \citenamefont {Sall}, \citenamefont {Mailly}, \citenamefont {Pham},
  \citenamefont {Zhang}, \citenamefont {Sisodia}, \citenamefont {Foerster}
  \emph {et~al.}}]{juge2021helium}%
  \BibitemOpen
  \bibfield  {author} {\bibinfo {author} {\bibfnamefont {R.}~\bibnamefont
  {Juge}}, \bibinfo {author} {\bibfnamefont {K.}~\bibnamefont {Bairagi}},
  \bibinfo {author} {\bibfnamefont {K.~G.}\ \bibnamefont {Rana}}, \bibinfo
  {author} {\bibfnamefont {J.}~\bibnamefont {Vogel}}, \bibinfo {author}
  {\bibfnamefont {M.}~\bibnamefont {Sall}}, \bibinfo {author} {\bibfnamefont
  {D.}~\bibnamefont {Mailly}}, \bibinfo {author} {\bibfnamefont {V.~T.}\
  \bibnamefont {Pham}}, \bibinfo {author} {\bibfnamefont {Q.}~\bibnamefont
  {Zhang}}, \bibinfo {author} {\bibfnamefont {N.}~\bibnamefont {Sisodia}},
  \bibinfo {author} {\bibfnamefont {M.}~\bibnamefont {Foerster}}, \emph
  {et~al.},\ }\bibfield  {title} {\bibinfo {title} {Helium ions put magnetic
  skyrmions on the track},\ }\href@noop {} {\bibfield  {journal} {\bibinfo
  {journal} {Nano Letters}\ }\textbf {\bibinfo {volume} {21}},\ \bibinfo
  {pages} {2989} (\bibinfo {year} {2021})}\BibitemShut {NoStop}%
\bibitem [{\citenamefont {Kern}\ \emph
  {et~al.}(2022{\natexlab{a}})\citenamefont {Kern}, \citenamefont {Pfau},
  \citenamefont {Deinhart}, \citenamefont {Schneider}, \citenamefont {Klose},
  \citenamefont {Gerlinger}, \citenamefont {Wittrock}, \citenamefont {Engel},
  \citenamefont {Will}, \citenamefont {G{\"u}nther} \emph
  {et~al.}}]{kern2022deterministic}%
  \BibitemOpen
  \bibfield  {author} {\bibinfo {author} {\bibfnamefont {L.-M.}\ \bibnamefont
  {Kern}}, \bibinfo {author} {\bibfnamefont {B.}~\bibnamefont {Pfau}}, \bibinfo
  {author} {\bibfnamefont {V.}~\bibnamefont {Deinhart}}, \bibinfo {author}
  {\bibfnamefont {M.}~\bibnamefont {Schneider}}, \bibinfo {author}
  {\bibfnamefont {C.}~\bibnamefont {Klose}}, \bibinfo {author} {\bibfnamefont
  {K.}~\bibnamefont {Gerlinger}}, \bibinfo {author} {\bibfnamefont
  {S.}~\bibnamefont {Wittrock}}, \bibinfo {author} {\bibfnamefont
  {D.}~\bibnamefont {Engel}}, \bibinfo {author} {\bibfnamefont
  {I.}~\bibnamefont {Will}}, \bibinfo {author} {\bibfnamefont {C.~M.}\
  \bibnamefont {G{\"u}nther}}, \emph {et~al.},\ }\bibfield  {title} {\bibinfo
  {title} {Deterministic generation and guided motion of magnetic skyrmions by
  focused {H}e+-ion irradiation},\ }\href@noop {} {\bibfield  {journal}
  {\bibinfo  {journal} {Nano Letters}\ }\textbf {\bibinfo {volume} {22}},\
  \bibinfo {pages} {4028} (\bibinfo {year} {2022}{\natexlab{a}})}\BibitemShut
  {NoStop}%
\bibitem [{\citenamefont {Kern}\ \emph
  {et~al.}(2022{\natexlab{b}})\citenamefont {Kern}, \citenamefont {Pfau},
  \citenamefont {Schneider}, \citenamefont {Gerlinger}, \citenamefont
  {Deinhart}, \citenamefont {Wittrock}, \citenamefont {Sidiropoulos},
  \citenamefont {Engel}, \citenamefont {Will}, \citenamefont {G{\"u}nther}
  \emph {et~al.}}]{kern2022tailoring}%
  \BibitemOpen
  \bibfield  {author} {\bibinfo {author} {\bibfnamefont {L.-M.}\ \bibnamefont
  {Kern}}, \bibinfo {author} {\bibfnamefont {B.}~\bibnamefont {Pfau}}, \bibinfo
  {author} {\bibfnamefont {M.}~\bibnamefont {Schneider}}, \bibinfo {author}
  {\bibfnamefont {K.}~\bibnamefont {Gerlinger}}, \bibinfo {author}
  {\bibfnamefont {V.}~\bibnamefont {Deinhart}}, \bibinfo {author}
  {\bibfnamefont {S.}~\bibnamefont {Wittrock}}, \bibinfo {author}
  {\bibfnamefont {T.}~\bibnamefont {Sidiropoulos}}, \bibinfo {author}
  {\bibfnamefont {D.}~\bibnamefont {Engel}}, \bibinfo {author} {\bibfnamefont
  {I.}~\bibnamefont {Will}}, \bibinfo {author} {\bibfnamefont {C.~M.}\
  \bibnamefont {G{\"u}nther}}, \emph {et~al.},\ }\bibfield  {title} {\bibinfo
  {title} {Tailoring optical excitation to control magnetic skyrmion
  nucleation},\ }\href@noop {} {\bibfield  {journal} {\bibinfo  {journal}
  {Physical Review B}\ }\textbf {\bibinfo {volume} {106}},\ \bibinfo {pages}
  {054435} (\bibinfo {year} {2022}{\natexlab{b}})}\BibitemShut {NoStop}%
\bibitem [{\citenamefont {Ahrens}\ \emph {et~al.}(2022)\citenamefont {Ahrens},
  \citenamefont {Gnoli}, \citenamefont {Giuliano}, \citenamefont {Mendisch},
  \citenamefont {Kiechle}, \citenamefont {Riente},\ and\ \citenamefont
  {Becherer}}]{ahrens2022skyrmion}%
  \BibitemOpen
  \bibfield  {author} {\bibinfo {author} {\bibfnamefont {V.}~\bibnamefont
  {Ahrens}}, \bibinfo {author} {\bibfnamefont {L.}~\bibnamefont {Gnoli}},
  \bibinfo {author} {\bibfnamefont {D.}~\bibnamefont {Giuliano}}, \bibinfo
  {author} {\bibfnamefont {S.}~\bibnamefont {Mendisch}}, \bibinfo {author}
  {\bibfnamefont {M.}~\bibnamefont {Kiechle}}, \bibinfo {author} {\bibfnamefont
  {F.}~\bibnamefont {Riente}},\ and\ \bibinfo {author} {\bibfnamefont
  {M.}~\bibnamefont {Becherer}},\ }\bibfield  {title} {\bibinfo {title}
  {{Skyrmion velocities in FIB irradiated W/CoFeB/MgO thin films}},\
  }\href@noop {} {\bibfield  {journal} {\bibinfo  {journal} {AIP Advances}\
  }\textbf {\bibinfo {volume} {12}},\ \bibinfo {pages} {035325} (\bibinfo
  {year} {2022})}\BibitemShut {NoStop}%
\bibitem [{\citenamefont {Yu}\ \emph {et~al.}(2016)\citenamefont {Yu},
  \citenamefont {Upadhyaya}, \citenamefont {Li}, \citenamefont {Li},
  \citenamefont {Kim}, \citenamefont {Fan}, \citenamefont {Wong}, \citenamefont
  {Tserkovnyak}, \citenamefont {Amiri},\ and\ \citenamefont
  {Wang}}]{yu2016room}%
  \BibitemOpen
  \bibfield  {author} {\bibinfo {author} {\bibfnamefont {G.}~\bibnamefont
  {Yu}}, \bibinfo {author} {\bibfnamefont {P.}~\bibnamefont {Upadhyaya}},
  \bibinfo {author} {\bibfnamefont {X.}~\bibnamefont {Li}}, \bibinfo {author}
  {\bibfnamefont {W.}~\bibnamefont {Li}}, \bibinfo {author} {\bibfnamefont
  {S.~K.}\ \bibnamefont {Kim}}, \bibinfo {author} {\bibfnamefont
  {Y.}~\bibnamefont {Fan}}, \bibinfo {author} {\bibfnamefont {K.~L.}\
  \bibnamefont {Wong}}, \bibinfo {author} {\bibfnamefont {Y.}~\bibnamefont
  {Tserkovnyak}}, \bibinfo {author} {\bibfnamefont {P.~K.}\ \bibnamefont
  {Amiri}},\ and\ \bibinfo {author} {\bibfnamefont {K.~L.}\ \bibnamefont
  {Wang}},\ }\bibfield  {title} {\bibinfo {title} {Room-temperature creation
  and spin--orbit torque manipulation of skyrmions in thin films with
  engineered asymmetry},\ }\href@noop {} {\bibfield  {journal} {\bibinfo
  {journal} {Nano Letters}\ }\textbf {\bibinfo {volume} {16}},\ \bibinfo
  {pages} {1981} (\bibinfo {year} {2016})}\BibitemShut {NoStop}%
\bibitem [{\citenamefont {Gusev}\ \emph {et~al.}(2020)\citenamefont {Gusev},
  \citenamefont {Sadovnikov}, \citenamefont {Nikitov}, \citenamefont
  {Sapozhnikov},\ and\ \citenamefont {Udalov}}]{gusev2020manipulation}%
  \BibitemOpen
  \bibfield  {author} {\bibinfo {author} {\bibfnamefont {N.}~\bibnamefont
  {Gusev}}, \bibinfo {author} {\bibfnamefont {A.}~\bibnamefont {Sadovnikov}},
  \bibinfo {author} {\bibfnamefont {S.}~\bibnamefont {Nikitov}}, \bibinfo
  {author} {\bibfnamefont {M.}~\bibnamefont {Sapozhnikov}},\ and\ \bibinfo
  {author} {\bibfnamefont {O.}~\bibnamefont {Udalov}},\ }\bibfield  {title}
  {\bibinfo {title} {Manipulation of the {D}zyaloshinskii--{M}oriya interaction
  in {C}o/{P}t multilayers with strain},\ }\href@noop {} {\bibfield  {journal}
  {\bibinfo  {journal} {Physical Review Letters}\ }\textbf {\bibinfo {volume}
  {124}},\ \bibinfo {pages} {157202} (\bibinfo {year} {2020})}\BibitemShut
  {NoStop}%
\bibitem [{\citenamefont {Tomasello}\ \emph {et~al.}(2018)\citenamefont
  {Tomasello}, \citenamefont {Komineas}, \citenamefont {Siracusano},
  \citenamefont {Carpentieri},\ and\ \citenamefont
  {Finocchio}}]{tomasello2018chiral}%
  \BibitemOpen
  \bibfield  {author} {\bibinfo {author} {\bibfnamefont {R.}~\bibnamefont
  {Tomasello}}, \bibinfo {author} {\bibfnamefont {S.}~\bibnamefont {Komineas}},
  \bibinfo {author} {\bibfnamefont {G.}~\bibnamefont {Siracusano}}, \bibinfo
  {author} {\bibfnamefont {M.}~\bibnamefont {Carpentieri}},\ and\ \bibinfo
  {author} {\bibfnamefont {G.}~\bibnamefont {Finocchio}},\ }\bibfield  {title}
  {\bibinfo {title} {Chiral skyrmions in an anisotropy gradient},\ }\href@noop
  {} {\bibfield  {journal} {\bibinfo  {journal} {Physical Review B}\ }\textbf
  {\bibinfo {volume} {98}},\ \bibinfo {pages} {024421} (\bibinfo {year}
  {2018})}\BibitemShut {NoStop}%
\bibitem [{\citenamefont {Gorshkov}\ \emph {et~al.}(2022)\citenamefont
  {Gorshkov}, \citenamefont {Gorev}, \citenamefont {Sapozhnikov},\ and\
  \citenamefont {Udalov}}]{gorshkov2022dmi}%
  \BibitemOpen
  \bibfield  {author} {\bibinfo {author} {\bibfnamefont {I.~O.}\ \bibnamefont
  {Gorshkov}}, \bibinfo {author} {\bibfnamefont {R.~V.}\ \bibnamefont {Gorev}},
  \bibinfo {author} {\bibfnamefont {M.~V.}\ \bibnamefont {Sapozhnikov}},\ and\
  \bibinfo {author} {\bibfnamefont {O.~G.}\ \bibnamefont {Udalov}},\ }\bibfield
   {title} {\bibinfo {title} {Dmi-gradient-driven skyrmion motion},\
  }\href@noop {} {\bibfield  {journal} {\bibinfo  {journal} {ACS Applied
  Electronic Materials}\ }\textbf {\bibinfo {volume} {4}},\ \bibinfo {pages}
  {3205} (\bibinfo {year} {2022})}\BibitemShut {NoStop}%
\bibitem [{\citenamefont {Vansteenkiste}\ and\ \citenamefont {Van~de
  Wiele}(2011)}]{vansteenkiste2011mumax}%
  \BibitemOpen
  \bibfield  {author} {\bibinfo {author} {\bibfnamefont {A.}~\bibnamefont
  {Vansteenkiste}}\ and\ \bibinfo {author} {\bibfnamefont {B.}~\bibnamefont
  {Van~de Wiele}},\ }\bibfield  {title} {\bibinfo {title} {Mumax: A new
  high-performance micromagnetic simulation tool},\ }\href@noop {} {\bibfield
  {journal} {\bibinfo  {journal} {Journal of Magnetism and Magnetic Materials}\
  }\textbf {\bibinfo {volume} {323}},\ \bibinfo {pages} {2585} (\bibinfo {year}
  {2011})}\BibitemShut {NoStop}%
\bibitem [{\citenamefont {Vansteenkiste}\ \emph {et~al.}(2014)\citenamefont
  {Vansteenkiste}, \citenamefont {Leliaert}, \citenamefont {Dvornik},
  \citenamefont {Helsen}, \citenamefont {Garcia-Sanchez},\ and\ \citenamefont
  {Van~Waeyenberge}}]{vansteenkiste2014design}%
  \BibitemOpen
  \bibfield  {author} {\bibinfo {author} {\bibfnamefont {A.}~\bibnamefont
  {Vansteenkiste}}, \bibinfo {author} {\bibfnamefont {J.}~\bibnamefont
  {Leliaert}}, \bibinfo {author} {\bibfnamefont {M.}~\bibnamefont {Dvornik}},
  \bibinfo {author} {\bibfnamefont {M.}~\bibnamefont {Helsen}}, \bibinfo
  {author} {\bibfnamefont {F.}~\bibnamefont {Garcia-Sanchez}},\ and\ \bibinfo
  {author} {\bibfnamefont {B.}~\bibnamefont {Van~Waeyenberge}},\ }\bibfield
  {title} {\bibinfo {title} {The design and verification of mumax3},\
  }\href@noop {} {\bibfield  {journal} {\bibinfo  {journal} {AIP Advances}\
  }\textbf {\bibinfo {volume} {4}},\ \bibinfo {pages} {107133} (\bibinfo {year}
  {2014})}\BibitemShut {NoStop}%
\bibitem [{\citenamefont {Landau}\ and\ \citenamefont
  {Lifshitz}(1992)}]{landau1992theory}%
  \BibitemOpen
  \bibfield  {author} {\bibinfo {author} {\bibfnamefont {L.}~\bibnamefont
  {Landau}}\ and\ \bibinfo {author} {\bibfnamefont {E.}~\bibnamefont
  {Lifshitz}},\ }\bibfield  {title} {\bibinfo {title} {On the theory of the
  dispersion of magnetic permeability in ferromagnetic bodies},\ }in\
  \href@noop {} {\emph {\bibinfo {booktitle} {Perspectives in Theoretical
  Physics}}}\ (\bibinfo  {publisher} {Elsevier},\ \bibinfo {year} {1992})\ pp.\
  \bibinfo {pages} {51--65}\BibitemShut {NoStop}%
\bibitem [{\citenamefont {Sampaio}\ \emph {et~al.}(2013)\citenamefont
  {Sampaio}, \citenamefont {Cros}, \citenamefont {Rohart}, \citenamefont
  {Thiaville},\ and\ \citenamefont {Fert}}]{sampaio2013nucleation}%
  \BibitemOpen
  \bibfield  {author} {\bibinfo {author} {\bibfnamefont {J.}~\bibnamefont
  {Sampaio}}, \bibinfo {author} {\bibfnamefont {V.}~\bibnamefont {Cros}},
  \bibinfo {author} {\bibfnamefont {S.}~\bibnamefont {Rohart}}, \bibinfo
  {author} {\bibfnamefont {A.}~\bibnamefont {Thiaville}},\ and\ \bibinfo
  {author} {\bibfnamefont {A.}~\bibnamefont {Fert}},\ }\bibfield  {title}
  {\bibinfo {title} {Nucleation, stability and current-induced motion of
  isolated magnetic skyrmions in nanostructures},\ }\href@noop {} {\bibfield
  {journal} {\bibinfo  {journal} {Nature Nanotechnology}\ }\textbf {\bibinfo
  {volume} {8}},\ \bibinfo {pages} {839} (\bibinfo {year} {2013})}\BibitemShut
  {NoStop}%
\bibitem [{\citenamefont {Romming}\ \emph {et~al.}(2015)\citenamefont
  {Romming}, \citenamefont {Kubetzka}, \citenamefont {Hanneken}, \citenamefont
  {von Bergmann},\ and\ \citenamefont {Wiesendanger}}]{romming2015field}%
  \BibitemOpen
  \bibfield  {author} {\bibinfo {author} {\bibfnamefont {N.}~\bibnamefont
  {Romming}}, \bibinfo {author} {\bibfnamefont {A.}~\bibnamefont {Kubetzka}},
  \bibinfo {author} {\bibfnamefont {C.}~\bibnamefont {Hanneken}}, \bibinfo
  {author} {\bibfnamefont {K.}~\bibnamefont {von Bergmann}},\ and\ \bibinfo
  {author} {\bibfnamefont {R.}~\bibnamefont {Wiesendanger}},\ }\bibfield
  {title} {\bibinfo {title} {Field-dependent size and shape of single magnetic
  skyrmions},\ }\href@noop {} {\bibfield  {journal} {\bibinfo  {journal}
  {Physical Review Letters}\ }\textbf {\bibinfo {volume} {114}},\ \bibinfo
  {pages} {177203} (\bibinfo {year} {2015})}\BibitemShut {NoStop}%
\bibitem [{\citenamefont {B{\"u}ttner}\ \emph {et~al.}(2018)\citenamefont
  {B{\"u}ttner}, \citenamefont {Lemesh},\ and\ \citenamefont
  {Beach}}]{buttner2018theory}%
  \BibitemOpen
  \bibfield  {author} {\bibinfo {author} {\bibfnamefont {F.}~\bibnamefont
  {B{\"u}ttner}}, \bibinfo {author} {\bibfnamefont {I.}~\bibnamefont
  {Lemesh}},\ and\ \bibinfo {author} {\bibfnamefont {G.~S.}\ \bibnamefont
  {Beach}},\ }\bibfield  {title} {\bibinfo {title} {Theory of isolated magnetic
  skyrmions: From fundamentals to room temperature applications},\ }\href@noop
  {} {\bibfield  {journal} {\bibinfo  {journal} {Scientific Reports}\ }\textbf
  {\bibinfo {volume} {8}},\ \bibinfo {pages} {4464} (\bibinfo {year}
  {2018})}\BibitemShut {NoStop}%
\bibitem [{\citenamefont {Wang}\ \emph {et~al.}(2018)\citenamefont {Wang},
  \citenamefont {Yuan},\ and\ \citenamefont {Wang}}]{wang2018theory}%
  \BibitemOpen
  \bibfield  {author} {\bibinfo {author} {\bibfnamefont {X.}~\bibnamefont
  {Wang}}, \bibinfo {author} {\bibfnamefont {H.}~\bibnamefont {Yuan}},\ and\
  \bibinfo {author} {\bibfnamefont {X.}~\bibnamefont {Wang}},\ }\bibfield
  {title} {\bibinfo {title} {A theory on skyrmion size},\ }\href@noop {}
  {\bibfield  {journal} {\bibinfo  {journal} {Communications Physics}\ }\textbf
  {\bibinfo {volume} {1}},\ \bibinfo {pages} {31} (\bibinfo {year}
  {2018})}\BibitemShut {NoStop}%
\bibitem [{\citenamefont {Sch{\"u}tte}\ \emph {et~al.}(2014)\citenamefont
  {Sch{\"u}tte}, \citenamefont {Iwasaki}, \citenamefont {Rosch},\ and\
  \citenamefont {Nagaosa}}]{schutte2014inertia}%
  \BibitemOpen
  \bibfield  {author} {\bibinfo {author} {\bibfnamefont {C.}~\bibnamefont
  {Sch{\"u}tte}}, \bibinfo {author} {\bibfnamefont {J.}~\bibnamefont
  {Iwasaki}}, \bibinfo {author} {\bibfnamefont {A.}~\bibnamefont {Rosch}},\
  and\ \bibinfo {author} {\bibfnamefont {N.}~\bibnamefont {Nagaosa}},\
  }\bibfield  {title} {\bibinfo {title} {Inertia, diffusion, and dynamics of a
  driven skyrmion},\ }\href@noop {} {\bibfield  {journal} {\bibinfo  {journal}
  {Physical Review B}\ }\textbf {\bibinfo {volume} {90}},\ \bibinfo {pages}
  {174434} (\bibinfo {year} {2014})}\BibitemShut {NoStop}%
\bibitem [{\citenamefont {B{\"u}ttner}\ \emph {et~al.}(2015)\citenamefont
  {B{\"u}ttner}, \citenamefont {Moutafis}, \citenamefont {Schneider},
  \citenamefont {Kr{\"u}ger}, \citenamefont {G{\"u}nther}, \citenamefont
  {Geilhufe}, \citenamefont {Schmising}, \citenamefont {Mohanty}, \citenamefont
  {Pfau}, \citenamefont {Schaffert} \emph {et~al.}}]{buttner2015dynamics}%
  \BibitemOpen
  \bibfield  {author} {\bibinfo {author} {\bibfnamefont {F.}~\bibnamefont
  {B{\"u}ttner}}, \bibinfo {author} {\bibfnamefont {C.}~\bibnamefont
  {Moutafis}}, \bibinfo {author} {\bibfnamefont {M.}~\bibnamefont {Schneider}},
  \bibinfo {author} {\bibfnamefont {B.}~\bibnamefont {Kr{\"u}ger}}, \bibinfo
  {author} {\bibfnamefont {C.}~\bibnamefont {G{\"u}nther}}, \bibinfo {author}
  {\bibfnamefont {J.}~\bibnamefont {Geilhufe}}, \bibinfo {author}
  {\bibfnamefont {C.~v.~K.}\ \bibnamefont {Schmising}}, \bibinfo {author}
  {\bibfnamefont {J.}~\bibnamefont {Mohanty}}, \bibinfo {author} {\bibfnamefont
  {B.}~\bibnamefont {Pfau}}, \bibinfo {author} {\bibfnamefont {S.}~\bibnamefont
  {Schaffert}}, \emph {et~al.},\ }\bibfield  {title} {\bibinfo {title}
  {Dynamics and inertia of skyrmionic spin structures},\ }\href@noop {}
  {\bibfield  {journal} {\bibinfo  {journal} {Nature Physics}\ }\textbf
  {\bibinfo {volume} {11}},\ \bibinfo {pages} {225} (\bibinfo {year}
  {2015})}\BibitemShut {NoStop}%
\end{thebibliography}%

\end{document}